\documentclass[structabstract]{aa}
\usepackage{graphicx}
\graphicspath{{figures/}}
\usepackage{subfig}
\usepackage{txfonts}
\usepackage{natbib}
\usepackage{url}
\usepackage{lscape}
\usepackage{rotating}
\usepackage{supertabular}
\usepackage{longtable}

\begin{document}
\title{The relation between $^{13}$CO~\textit{J}\,=\,2\,--\,1 line width in molecular clouds and bolometric luminosity of associated IRAS sources}
\authorrunning{Ke Wang et al.}
\titlerunning{Linking massive YSOs to their parent clouds}

\author{Ke Wang\inst{1} \and Yuefang
Wu\inst{1} \and Liang Ran\inst{1,2} \and Wentao Yu \inst{3} \and
Martin Miller\inst{4}}

\institute{Department of Astronomy, School of Physics, Peking
University, Beijing 100871, China\\ \email{kwang@cfa.harvard.edu yfwu@bac.pku.edu.cn} \and
Department of Atmospheric Sciences, School of Physics, Peking
University, Beijing 100871, China \and
Institut f\"{u}r Anorganische Chemie, Universit\"{a}t Bonn, R\"{o}mer St.\,164, D-53117 Bonn, Germany
\and I. Physikal. Institut, Universit\"{a}t zu K\"{o}ln,
Z\"{u}lpicher St.\,77, D-50937 K\"{o}ln, Germany}

   \date{Received ?; accepted ?}

% \abstract{}{}{}{}{}
% 5 {} token are mandatory

\abstract
  {}
{We search for evidence of a relation between properties of young
stellar objects (YSOs) and their parent molecular clouds to
understand the initial conditions of high-mass star formation.}
{A sample of 135 sources was selected from the Infrared Astronomical
Satellite (\textit{IRAS}) Point Source Catalog, on the basis of
their red color to enhance the possibility of discovering young
sources. Using the K\"olner Observatorium f\"ur SubMillimeter
Astronomie (KOSMA) 3-m telescope, a single-point survey in
$^{13}$CO~\textit{J}\,=\,2\,--\,1 was carried out for the entire
sample, and 14 sources were mapped further. Archival mid-infrared
(MIR) data were compared with the $^{13}$CO emissions to identify
evolutionary stages of the sources. A $^{13}$CO observed sample was
assembled to investigate the correlation between $^{13}$CO line
width of the clouds and the luminosity of the associated YSOs.}
{We identified 98 sources suitable for star formation analyses for
which relevant parameters were calculated. We detected 18 cores from
14 mapped sources, which were identified with eight pre-UC H\,{\sc ii}
regions and one UC H\,{\sc ii} region, two high-mass cores earlier
than pre-UC H\,{\sc ii} phase, four possible star forming clusters,
and three sourceless cores. By compiling a large (360 sources)
$^{13}$CO observed sample, a good correlation was found between the
$^{13}$CO line width of the clouds and the bolometric luminosity of
the associated YSOs, which can be fitted as a power law,
$\lg\,(\Delta V_{13}/\,\rm{km\,s}^{-1}) = (-0.023\pm
0.044)+(0.135\pm 0.012)\lg\,(L_{\rm bol}/L_{\odot})$. Results show
that luminous ($>10^3\,L_{\odot}$) YSOs tend to be associated with
both more massive and more turbulent ($\Delta
V_{13}>2\,\rm{km\,s}^{-1}$) molecular cloud structures.}
  {}

\keywords{stars: formation --- ISM: clouds --- ISM: molecules --- ISM: kinematic and
dynamics}

\maketitle  %warning: no space in abstract!
%
%________________________________________________________________

\section{Introduction}

The past decade has witnessed significant progress in the study of
high-mass star formation. Observations at millimeter and
submillimeter wavelengths \citep{1998ApJ...505L.151Z,
2002A&A...383..892B, 2002ApJ...568..754K, 2005IAUS..227..135Z,
2007prpl.conf..197C} suggest that massive proto B stars can form by
disk mediated accretion, which is similar to the scenario that
produces low-mass stars. However, most of the studies focus on
relatively evolved stages, when the central star has already formed
and hydrogen burning has begun, characterized by surrounding ultra
compact (UC) H\,{\sc ii} regions and strong emission from complex
molecules \citep{2002ARA&A..40...27C}. In contrast, the extremely
early stages are poorly understood to date. In particular, knowledge
to evolutionary stages prior to the onset of H\,{\sc ii} regions are
crucial to understanding the initial conditions of high-mass star
formation.

It is known that stars are formed in molecular clouds. Therefore,
the relation between forming stars and parent clouds is important to
understand the formation process and the properties of the eventual
stars. On galaxy scales, star formation activities are usually
described by the so-called Schmidt law, which relates the star
formation rate (SFR) to the surface density of gas: $\Sigma_{\rm
SFR} \propto \Sigma _{\rm gas}^{N}$, where the index $N=1-2$
\citep{1959ApJ...129..243S, 1998ApJ...498..541K,
2004ApJ...606..271G}. Studies of Galactic dense cores have shown
that this relation may be universal and can be connected to Galactic
star formation \citep{2005ApJ...635L.173W}.
\cite{1981MNRAS.194..809L} studied the turbulence in star forming
clouds and found a strong correlation between the internal velocity
dispersion $\sigma$ of the region and its size $L$:
$\sigma(\rm{km\,s}^{-1}) \propto $\,$L(\rm{pc})^{0.38}$. This
relation, also called the Larson law, is valid for low-mass cores
but is found to be break down in high-mass cores $\gtrsim
10^3M_{\odot}$ \citep{1995ApJ...446..665C, 1997ApJ...476..730P,
2008MNRAS.391..869G}. This is indicative of the different status of
turbulence in low- and high-mass cores. The breakdown of the Larson
law can be interpreted as evidence of widespread supersonic
turbulence in high-mass cores, in contrast to subsonic turbulent
low-mass cores \citep{1997ApJ...476..730P}. A molecular line width
is an observational indicator of turbulence in clouds, and
bolometric luminosity is an indicator of forming stars. Any relation
between these quantities may help us to understand the initial star
forming process.

Here we report results from a $^{13}$CO~\textit{J}\,=\,2\,--\,1
survey towards 135 \textit{IRAS} sources using the KOSMA 3-m
telescope. To search for high-mass star forming regions in their
early stages, we select a sample on the basis of their red
\textit{IRAS} color to enhance the possibility of finding young
sources. We present the primary results and investigate the relation
between line width in molecular clouds and bolometric luminosity of
associated infrared sources. We describe our sample selection in
Sect.\,\ref{sect:sample} and observations in Sect.\,\ref{sect:obs}.
In Sect.\,\ref{sect:results} we present statistical results of the
single-point survey (Sect.\,\ref{subsect:survey}) and follow-up
mapping (Sect.\,\ref{subsect:map}). We discuss the $\Delta V - L$
relation as well as other relations in Sect.\,\ref{sect:discussion},
and summarize the paper in Sect.\,\ref{sect:summary}.

%__________________________________________________________________
\section{Sample} \label{sect:sample}

We selected the sample from the Infrared Astronomical Satellite
(\textit{IRAS}) Point Source Catalog (PSC,
\citealt{1988iras....1.....B}) version 2.1 according to our
developed color criteria \citep{2003ChPhL..20.1409W}, namely:

\indent (a) $f_{100\mu m}<$~500~Jy, lg$(f_{25\mu m}/f_{12\mu
m})\geqslant$~0.7, lg$(f_{60\mu m}/f_{12\mu m})\geqslant$~1.4, where $f_{\lambda}$ is the flux
density;\\
\indent (b) lack of 6~cm radio continuum radiation to exclude potential H\,{\sc ii} associations;\\
\indent (c) declination $\delta > -20^{\circ}$, so that targets are
accessible to the telescope KOSMA.

Criterion (a) was chosen so that the sample sources were redder and
possibly fainter, thus may be younger than those selected based on
traditional \cite{1989ApJ...340..265W} color criteria. Criterion (b)
helps to exclude any known H\,{\sc ii} regions brighter than current
detection limit. Therefore, the sample should represent extremely
young stellar objects (YSOs), mostly at evolutionary stages earlier
than the UC H\,{\sc ii} phase. The 6~cm radio continuum data was
extracted from three surveys: (1) $0^{\circ}<\delta<75^{\circ}$
4.85\,GHz radio continuum survey completed by
\citet{1991ApJS...75.1011G} with the 91-m NRAO telescope; (2)
$-29^{\circ}<\delta<9.5^{\circ}$ 4.85\,GHz radio continuum survey
led by \citet{1994ApJS...90..179G} with the 64-m Parkes telescope;
and (3) $-9.5^{\circ}<\delta<10^{\circ}$ 4.85\,GHz radio continuum
survey led by \citet{1995ApJS...97..347G} with the 64-m Parkes
telescope.

Criteria (a) and (c) lead to 500 sources being selected from the
PSC, which contains 245,889 sources. However, only 135 sources were
observed because of limited observing time and after applying
criterion (b). These sources represent the sample reported in this
paper. The sample sources are concentrated across the Galactic plane
and cover a wide range of longitude, $10^{\circ}<l<230^{\circ}$.

\section{Observations} \label{sect:obs}
A single-point survey in $^{13}$CO~\textit{J}\,=\,2\,--\,1
(220.398\,GHz) was carried out from September 2002 to March 2003
using the K\"olner Observatorium f\"ur SubMillimeter Astronomie
(KOSMA\footnote{The KOSMA 3\,m radiotelescope at
Gornergrat-S\"ud Observatory is operated by the University of
Cologne and supported by special funding from the Land NRW. The
Observatory is administered by the Internationale Stiftung
Hochalpine Forschungsstationen Jungfraujoch und Gornergrat,
Bern.})\ 3-m telescope on Gornergrat near Zermatt in Switzerland.
All of the sample sources were surveyed in
$^{13}$CO~\textit{J}\,=\,2\,--\,1. About half of the sample sources
were also observed in $^{12}$CO~\textit{J}\,=\,2\,--\,1
(230.538\,GHz) and 14 of them were mapped in
\mbox{$^{13}$CO~\textit{J}\,=\,2\,--\,1.}

The beamwidth of the KOSMA at 230\,GHz was 130$^{\prime\prime}$. The
pointing accuracy was superior to 10$^{\prime\prime}$. The telescope
was equipped with a dual-channel SIS receiver, which had a noise
temperature of 150\,K. A high resolution spectrometer with 2048
channels was employed and the spectral resolution was 165.5\,KHz,
giving a velocity resolution of 0.22\,$\rm{km\,s}^{-1}$. The main
beam temperature ($T_{\rm{mb}}$) had been corrected for the effects
of Earth's atmosphere, antenna cover loss, radiation loss, and
forward spillover and scattering efficiency (92$\%$). From the
calibrated Jupiter observations, the main beam efficiency
$\eta_{mb}$ was estimated as 68$\%$ during our observation.
On-the-fly mode was adopted during mapping, with a mapping step of
60$^{\prime\prime}$. Most maps were extended until the line
intensity decreased to half of the maximum value or even lower. The
GILDAS\footnote{~available at
\url{http://www.iram.fr/IRAMFR/GILDAS}} software package
(CLASS/GREG/SIC) was used for the data reduction
\citep{2000ASPC..217..299G}.

\section{Results} \label{sect:results}
\subsection{Survey} \label{subsect:survey}

\begin{figure}
 \centering
 \includegraphics[width=9cm]{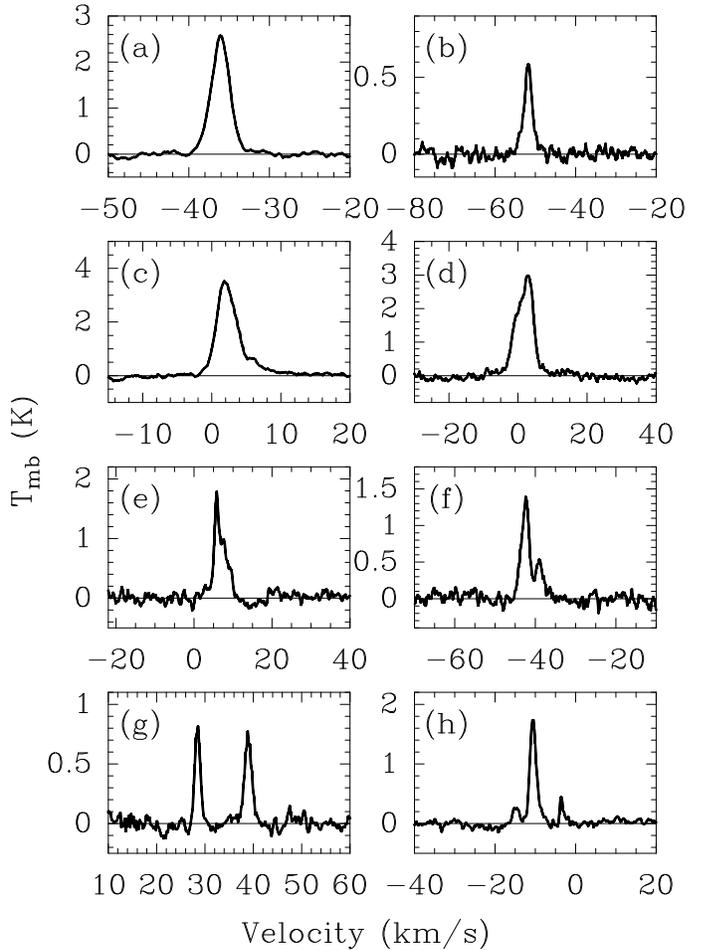}
\caption{Example spectra of $^{13}$CO~\textit{J}\,=\,2\,--\,1
towards the \textit{IRAS} sources given in the text.
\label{fig:spec}}
\end{figure}

Among the entire sample of 135 \textit{IRAS} sources, we identified
98 sources suitable for star formation analyses (another 37 sources
were excluded either because they had multiple components or bad
baselines, or failed to be detected), of which 60 have both
$^{13}$CO~\textit{J}\,=\,2\,--\,1  and
$^{12}$CO~\textit{J}\,=\,2\,--\,1  data. Figure~\ref{fig:spec} 
presents example spectra of $^{13}$CO~\textit{J}\,=\,2\,--\,1:
(a) \textit{IRAS} 00117+6412, a perfect Gaussian profile; (b)
\textit{IRAS} 02541+6208, a fairly narrow line width; (c)
\textit{IRAS} 06067+2138, a broad line with red wing, also seen in
\textit{J}\,=\,1\,--\,0 transition \citep{2003ChPhL..20.1409W}; (d)
\textit{IRAS} 20326+3757, a blue wing; (e) \textit{IRAS}
18278$-$0212, red asymmetry; (f) \textit{IRAS} 21379+5106, two
peaks; (g) \textit{IRAS} 19348+2229, two components; and (h)
\textit{IRAS} 02485+6902, multiple components.

Observed and derived parameters are listed in an online
Table~\ref{tab:survey98}, starting with \textit{IRAS} name and its
J2000 equatorial coordinates in Cols. (1) to (3). By Gaussian fit,
we obtain the observed parameters including main beam temperature
$T_{\rm{mb}}$, local standard of rest velocity $V_\mathrm{LSR13}$,
and $^{13}$CO~\textit{J}\,=\,2\,--\,1 line width (full width at
half-maximum) $\Delta V_{13}$ for each source, listed in Cols. (4)
to (7). When a line profile is obviously non-Gaussian, the
parameters are measured with a cursor (e.g.,
\citealt{2003ApJ...592L..79W}), and the velocity uncertainty is
given as the velocity resolution; when the line profile has
distinctive multiple components, only the strongest component is shown,
indicated by a character \textit{m} in corresponding $T_{\rm{mb}}$
columns.

The distance to most sources was unavailable in the literature. The
kinematic distances were calculated based on the radial velocity
$V_\mathrm{LSR13}$~and the velocity field of the outer Galaxy given
by \citet{1993A&A...275...67B}. When two kinematic distances were
available, we selected the closer one, except when the closer
distance is too small ($<$100\,pc). For 8 sources, however, no
reasonable distances could be calculated in this way and we assumed
that the distance to these sources is 1\,kpc. These are marked as *
in the distance Col. (8) of Table~\ref{tab:survey98}.

The bolometric luminosity was calculated based on the distances and
the \textit{IRAS} fluxes in four bands (12, 25, 60, 100\,$\mu m$),
following the formula given by \citet{1986A&A...169..281C}
\begin{eqnarray*}
L_{\rm{bol}} = 5.4D^{2}(f_{12\mu m}/0.79+f_{25\mu m}/2+f_{60\mu
m}/3.9+f_{100\mu m}/9.9)\,L_{\odot},
\end{eqnarray*}
where $D$ is the distance in kpc and $f_{\lambda}$ is the flux
density in Jansky. The uncertainties in luminosity originate in the
kinematic distances and the quality of the \textit{IRAS} source
fluxes. Most of the sample sources have high or moderate quality in
all the four bands. Twenty-one sources with upper limit fluxes in
one or two bands are marked as \textit{u} luminosity in Col. (9) of
Table~\ref{tab:survey98}.

Assuming local thermodynamic equilibrium (LTE) and that the
$^{13}$CO~\textit{J}\,=\,2\,--\,1 transition is optically thin
(\emph{i.e.} $\tau_{13}<1$), we derive excitation temperatures,
optical depth and column densities for $^{13}$CO, using radiation
transfer equation \citep{1991ApJ...374..540G}. Based on the
assumption of LTE, $^{13}$CO and $^{12}$CO share the same excitation
temperature $T_{\rm{ex}}$, which can be derived from the main beam
temperature of optically thick $^{12}$CO, $T_{\rm{mb12}}$. When
$\tau_{13}$$\,>1$, an optical depth correction factor $C_\tau =
\tau_{13}/(1-e^{-\tau_{13}})$ is multiplied by its corresponding
column density. The relative CO abundance [$^{12}$CO/H$_2$] is
estimated to extend from $2.5\times 10^{-5}$
\citep{1982ApJ...260..635R} to $10^{-4}$
\citep{1991ApJ...374..540G}, and we adopt the median value of
$6.25\times 10^{-5}$. Using the terrestrial [$^{12}$C/$^{13}$C]
ratio of 89, we adopt a value for [$^{13}$CO/H$_2$] of 7.0$\times
10^{-7}$ when computing the column density of H$_2$. These
parameters are listed in Cols. (10) to (13). References of former
works are given in the last Col. (14) of Table~\ref{tab:survey98}.

The distribution of $^{13}$CO~\textit{J}\,=\,2\,--\,1  line width of
this sample has a mean of 3.09\,$\rm{km\,s}^{-1}$~and a standard
deviation of 1.06\,$\rm{km\,s}^{-1}$. This line width is relatively
smaller than that of typical bright/red \textit{IRAS} sources
associated with water masers (3.5\,$\rm{km\,s}^{-1}$,
\citealt{2001A&A...380..665W}; note that this value was measured in
\textit{J}\,=\,1\,--\,0~transition), while significantly larger than
that of a molecular cloud hosting intermediate-mass star formation
activities ($\sim$2\,$\rm{km\,s}^{-1}$,
\citealt{2006A&A...451..539S}, averaged throughout the Perseus
cloud). The luminosities are distributed over a wide range, from
20\,$L_{\odot}$~to about $10^5$\,$L_{\odot}$, with a mean of
$10^4$\,$L_{\odot}$. The high dispersion of luminosities indicates
that these sources are embedded in very different environments. This
luminosity distribution is similar to the young \textit{'low'}
sources of \citealt{1996A&A...308..573M} (see their Fig.~6), in
agreement with the assumption that our sample group may be
relatively younger than that chosen by traditional color criteria.
The excitation temperature $T_\mathrm{ex}$~ranges from 4.4 to
22.5\,K, with an average of 9.7\,K. This suggests that very cold
gases surround the sample sources, colder than those
surrounding the luminous \textit{IRAS}  sources
\citep{2007ChJAA...7..331Z}. The $^{13}$CO column densities are
$(1.2-28.7)\,\times$$10^{15}\,\mathrm{cm}^{-2}$, with an average of
$6.2\,\times$$10^{15}\,\mathrm{cm}^{-2}$, while H$_2$ column
densities are $(1.7-40.8)\,\times$$10^{21}\,\mathrm{cm}^{-2}$, with
an average of $8.9\,\times$$10^{21}\,\mathrm{cm}^{-2}$. These
densities are roughly close to the critical value for gravitational
collapse \citep{1998masg.conf..101H}.

\vskip -0.03\textwidth
\setcounter{table}{1}
\begin{table*}
\begin{center}
\caption[]{Core Properties}
\begin{tabular}{lrc ccrrc}
\hline\hline\\
{Core} & {$R$} & {$\overline{\Delta V}_{13}$} &{$L_{\rm{bol}}$} & {Peak $n(\mathrm{H}_{2})$} & {$M_\mathrm{LTE}$ } & {$M_\mathrm{vir}$ } & {$\alpha ^*$}\\
{ } & {(pc)} & {($\rm{km\,s}^{-1}$)}  &{($10^3L_{\odot}$)}  & {($10^{3}\,\mathrm{cm}^{-3}$)} &{($M_{\odot}$)} & {($M_{\odot}$)} & {}\\
{(1)}     & {(2)}     & {(3)} & {(4)} & {(5)} & {(6)}  & {(7)} & {(8)}\\
\hline

00117+6412               &    0.57                &2.58                &4.29                &5.13                 &     1.8E+2         &4.8E+2        &2.7   \\
00557+5612               &    1.08                &2.05                &1.47$^{\rm u}$      &1.66                 &     6.6E+2         &5.7E+2        &0.9   \\
03101+5821               & $>$1.08                &2.52                &1.20                &1.76                 &  $>$5.8E+2         &8.6E+2        &1.5   \\
03260+3111$^{\rm a}$     & $>$0.23                &3.05                &0.29                &15.96                &  $>$1.6E+2         &2.6E+2        &1.6   \\
03260+3111NE$^{\rm s}$   &    0.11                &2.51                &...                 &15.96                &     8.0E+1         &8.8E+1        &1.1   \\
03414+3200               & $>$0.34                &1.92                &0.05$^{\rm u}$      &5.35                 &  $>$8.5E+1         &1.6E+2        &1.8   \\
05168+3634               & $>$2.41                &2.86                &17.13               &1.57                 &  $>$1.2E+4         &2.4E+3        &0.2   \\
05168+3634SW$^{\rm s}$   &    0.60$^{\rm m}$      &2.61                &...                 &1.57                 &     3.0E+3         &5.2E+2        &0.2   \\
06067+2138               &    0.55                &3.36$^{\rm b}$      &0.03$^{\rm u}$      &3.83 $^{\rm c}$      &     2.5E+2         &7.8E+2        &3.1   \\
06103+1523               &    1.97                &2.56                &9.49                &1.02                 &     2.8E+3         &1.6E+3        &0.6   \\
07024$-$1102             & $>$0.11$^{\rm m}$      &1.99                &0.57                &64.73$^{\rm c}$      &  $>$9.0E+0         &5.8E+1        &6.4   \\
20067+3415               &    1.76                &3.80$^{\rm b}$      &1.14                &1.14                 &     3.9E+3         &3.2E+3        &0.8   \\
20067+3415NE$^{\rm s}$   &    0.45$^{\rm m}$      &3.67                &...                 &1.14                 &     9.7E+2         &7.5E+2        &0.8   \\
20149+3913               &    0.43                &3.19                &0.30$^{\rm u}$      &2.62                 &     1.4E+3         &5.5E+2        &0.4   \\
20151+3911$^{\rm a}$     & $>$1.30                &3.83                &0.70$^{\rm u}$      &2.62                 &  $>$4.1E+3         &2.4E+3        &0.6   \\
21391+5802               &    0.32                &2.78                &0.26                &11.87$^{\rm c}$      &     1.3E+2         &3.1E+2        &2.4   \\
22198+6336               &    1.22                &2.21$^{\rm b}$      &1.53$^{\rm u}$      &2.39                 &     1.7E+3         &7.5E+2        &0.4   \\
22506+5944               &    1.31                &2.88                &6.83                &1.40                 &     1.0E+3         &1.4E+3        &1.4   \\
\hline
Average$^\dagger$        &    0.98                &2.80                &3.02                &4.95                 &     1.9E+3         &1.1E+3        &1.3   \\
\hline \label{tab:map14}
\end{tabular}
\end{center}
Note.--- $^{\rm a}$~not guide \textit{IRAS} source. Luminosity
calculated based on \textit{IRAS} fluxes and distance: 03260+3111 at
0.4\,kpc \citep{1998A&AS..132..211H} and 20151+3911 at 1.7\,kpc
\citep{2007A&A...476.1243M}. $^{\rm b}$~average spectrum fitted with
two Gaussian profiles, list the stronger one. $^{\rm c}$~no
$^{12}$CO data, density calculated by assuming reasonable
$T_\mathrm{ex}$ (see text). $^{\rm m}$~core's angular diameter
comparable to beamwidth, marginally resolved. $^{\rm s}$~sourceless
core. $^{\rm u}$~upper limit. $^\dagger$~average calculating does not
include values of marginally resolved cores, except for the line width column.
$^*$~$\alpha = M_\mathrm{vir}$/$M_\mathrm{LTE}$.
\end{table*}

\subsection{Mapping} \label{subsect:map}

\begin{figure*}
\centering \vskip -0.03\textwidth
\subfloat{\includegraphics[width=0.235\textwidth,height=0.3\textwidth,angle=-90]{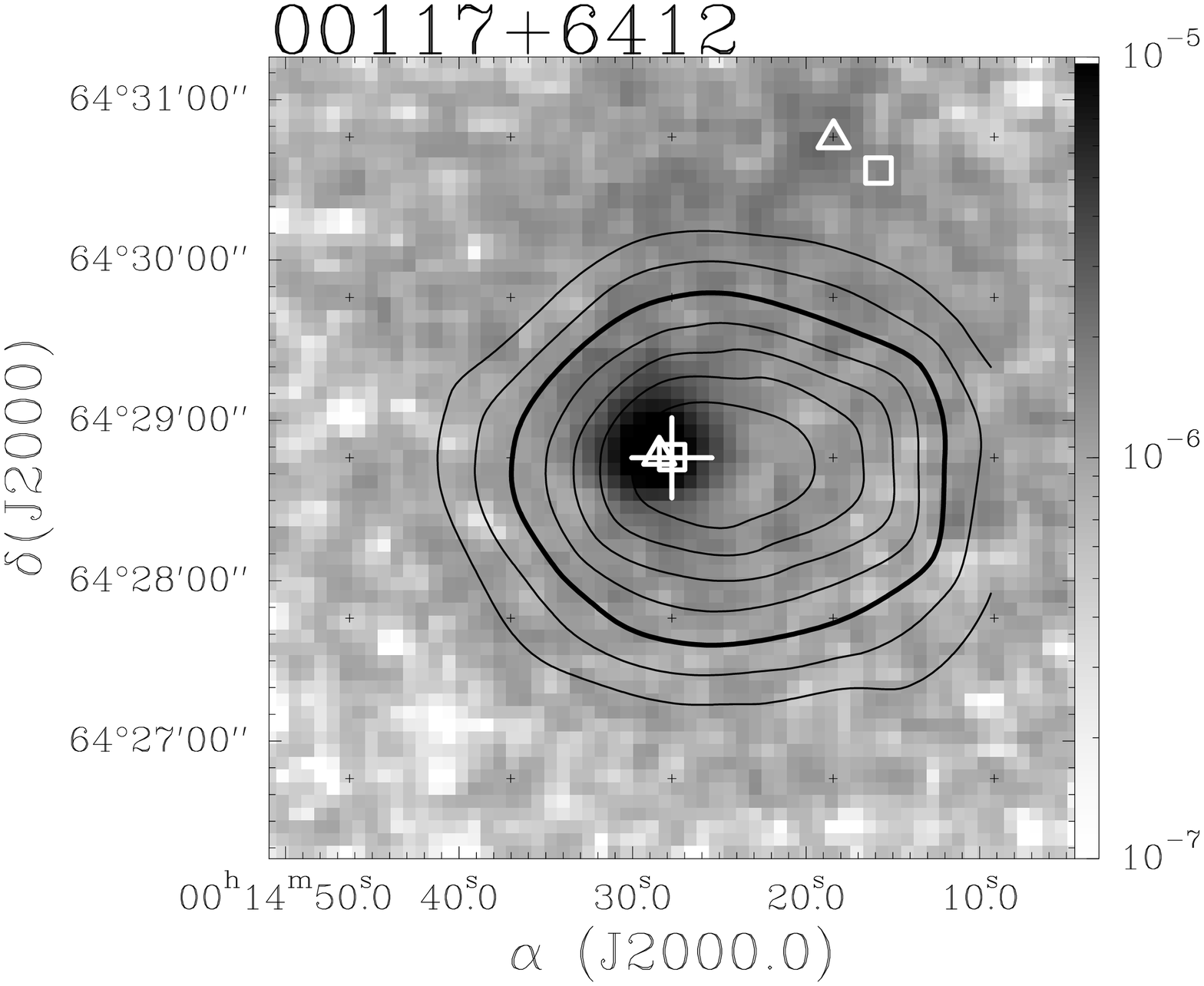}}
\subfloat{\includegraphics[width=0.235\textwidth,height=0.3\textwidth,angle=-90]{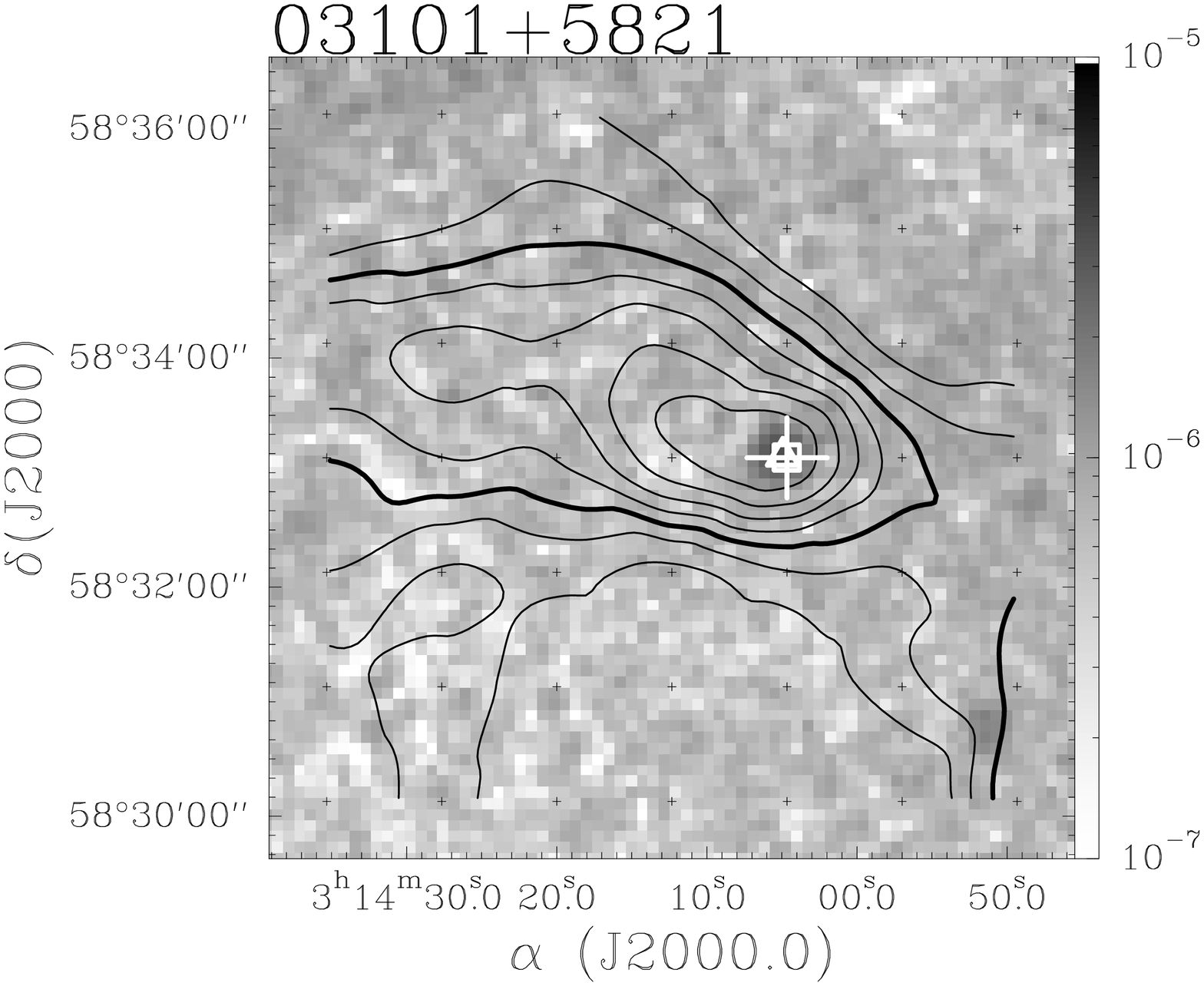}}
\subfloat{\includegraphics[width=0.235\textwidth,height=0.3\textwidth,angle=-90]{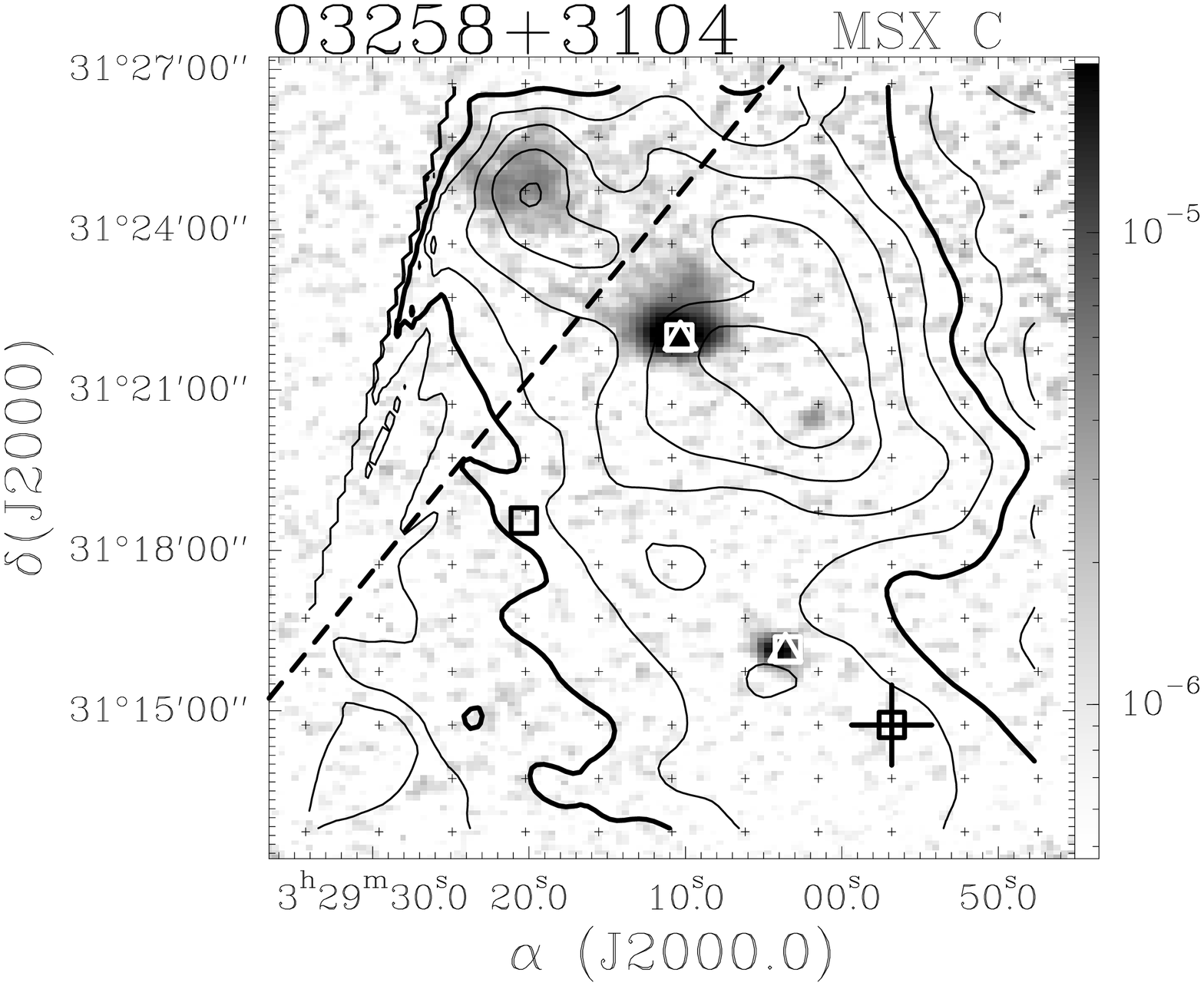}}\par
\vskip -0.01\textwidth
\subfloat{\includegraphics[width=0.235\textwidth,height=0.3\textwidth,angle=-90]{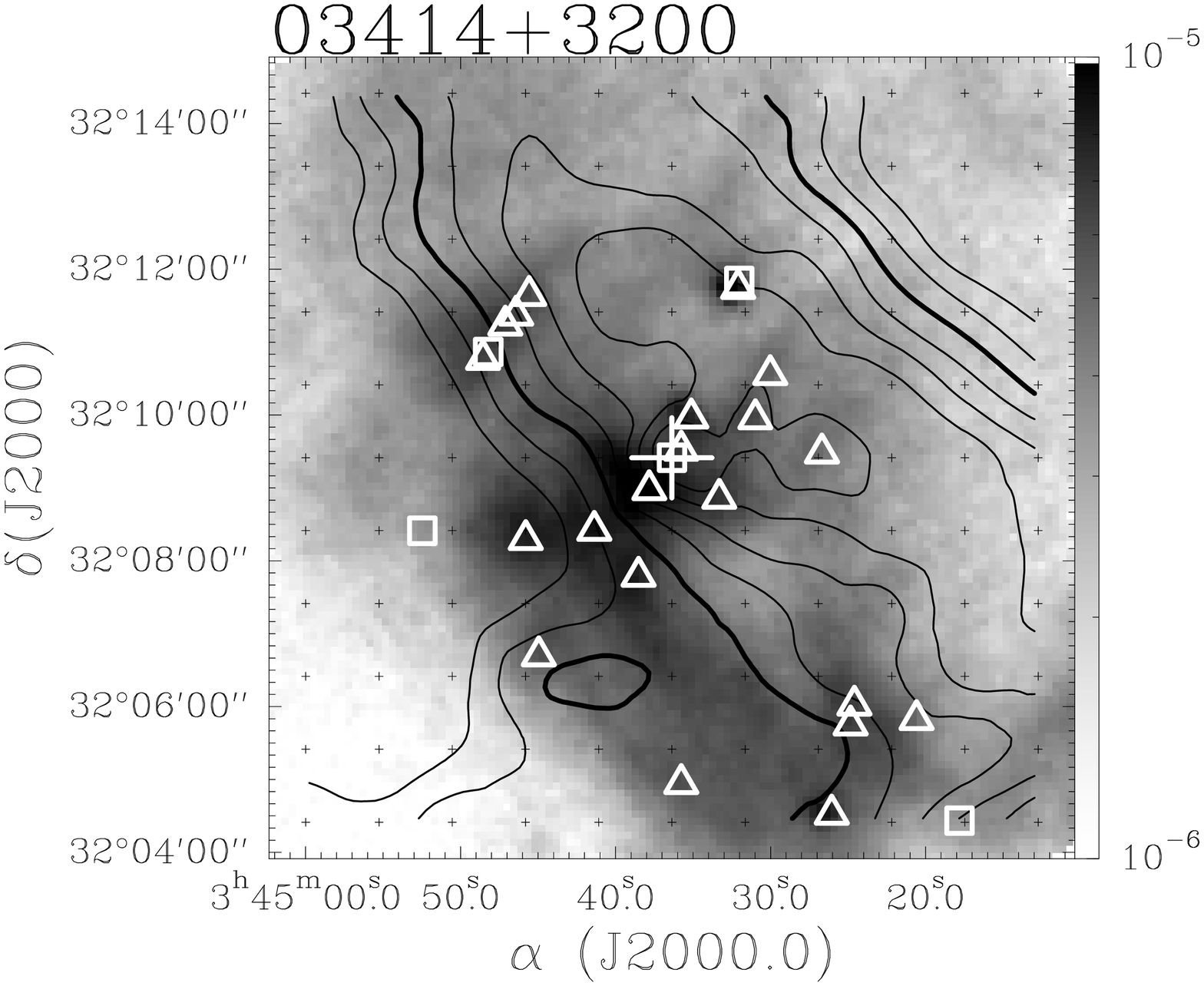}}
\subfloat{\includegraphics[width=0.235\textwidth,height=0.3\textwidth,angle=-90]{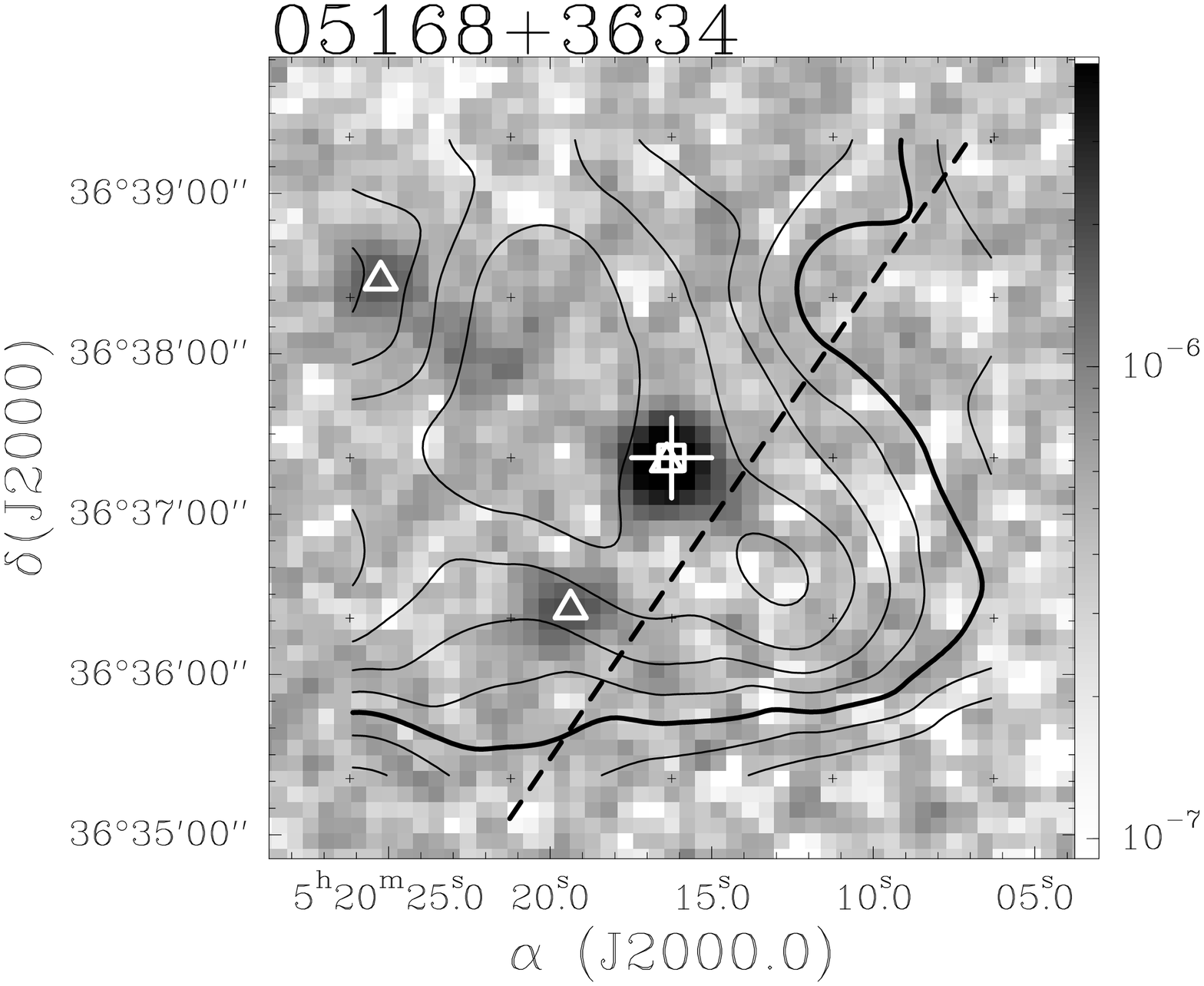}}
\subfloat{\includegraphics[width=0.235\textwidth,height=0.3\textwidth,angle=-90]{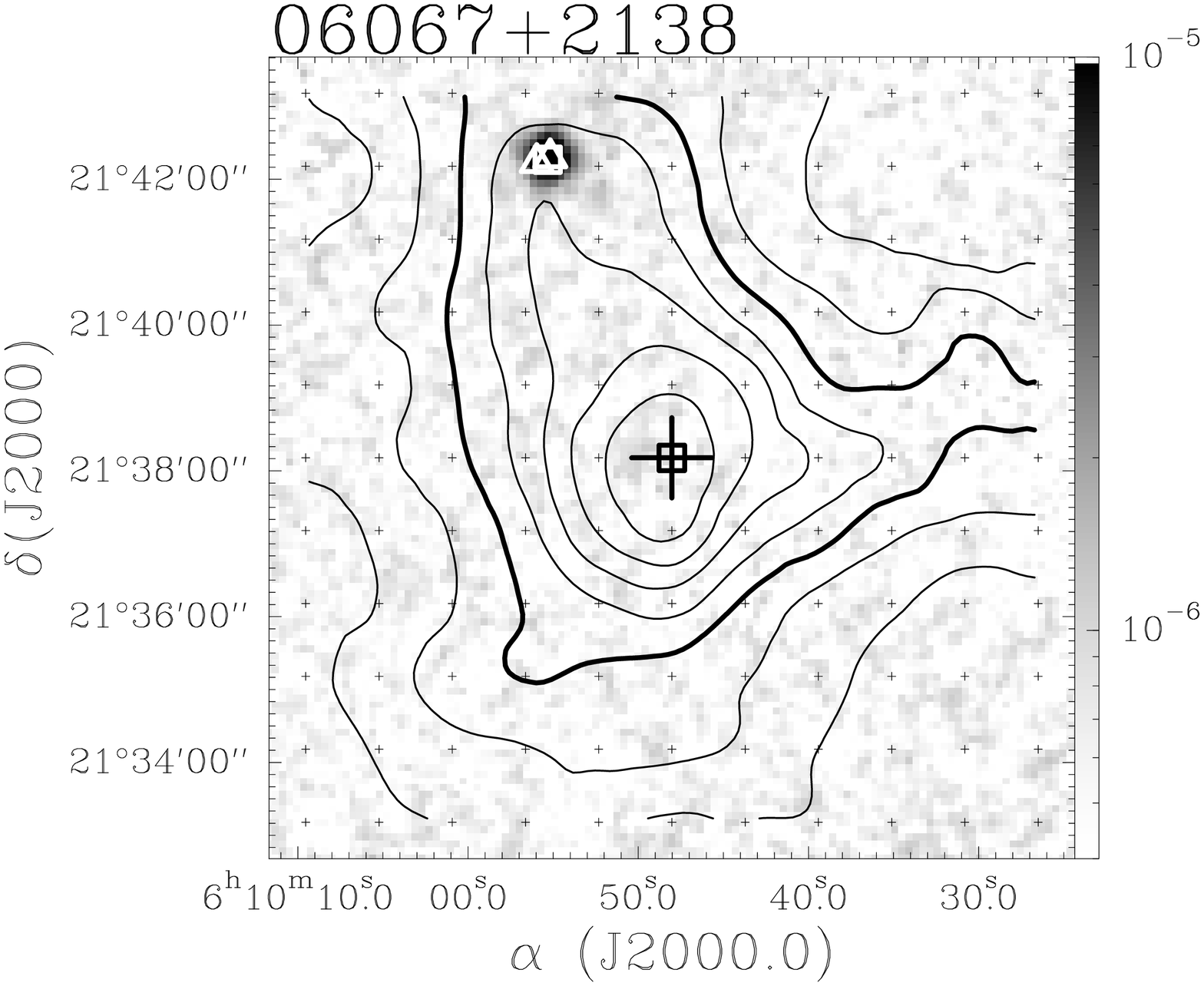}}
\vskip -0.01\textwidth
\subfloat{\includegraphics[width=0.235\textwidth,height=0.3\textwidth,angle=-90]{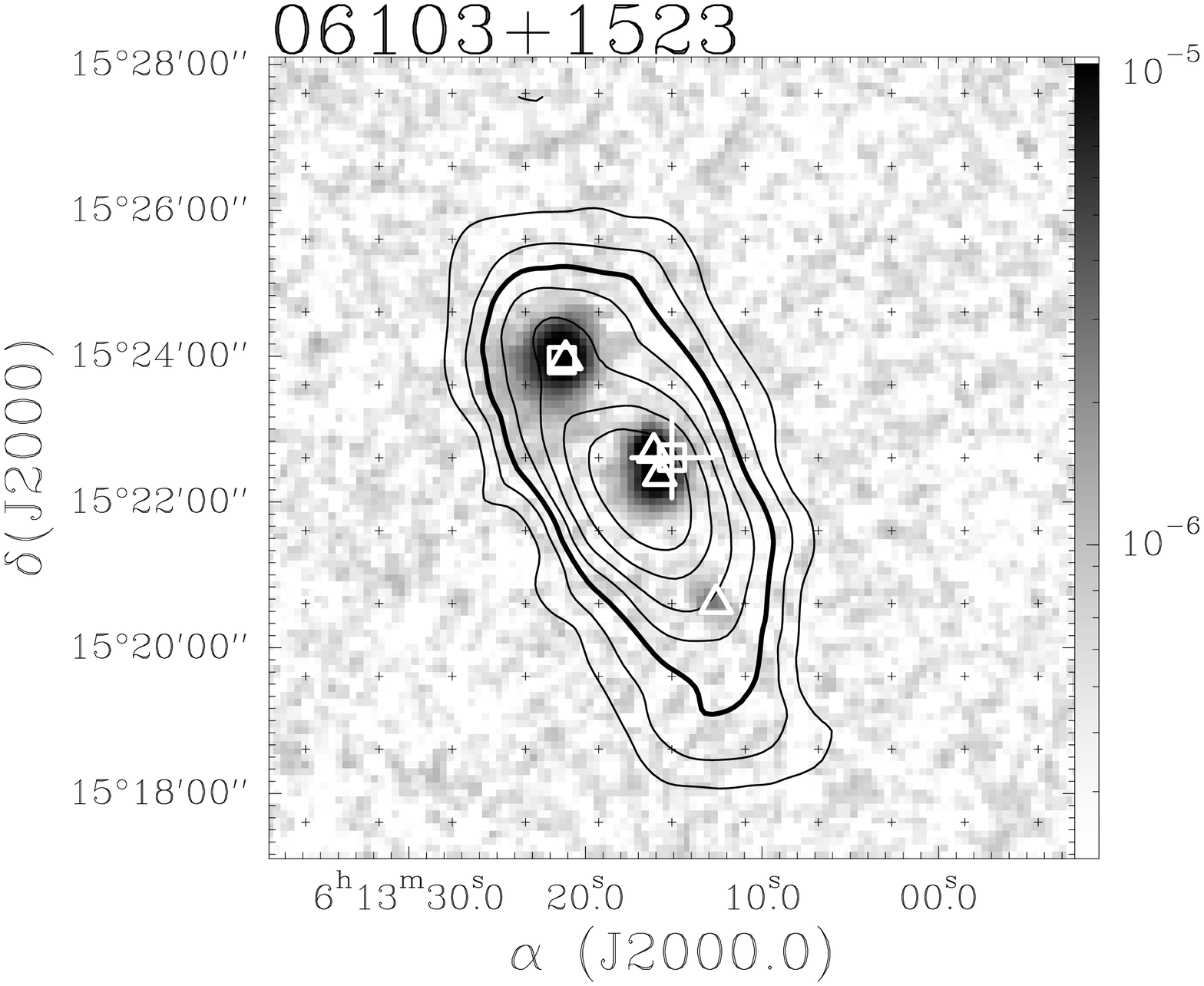}}
\subfloat{\includegraphics[width=0.235\textwidth,height=0.3\textwidth,angle=-90]{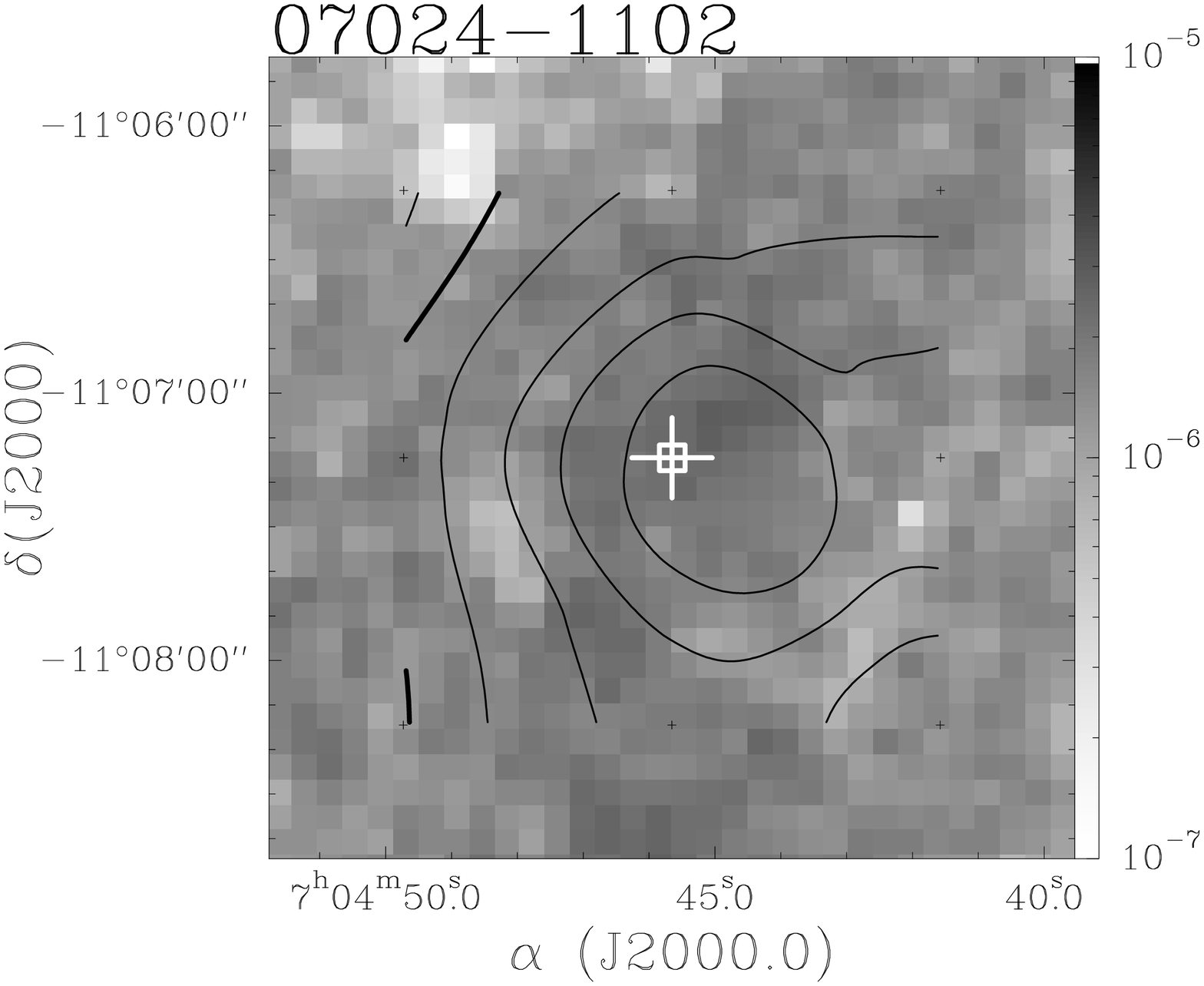}}
\subfloat{\includegraphics[width=0.235\textwidth,height=0.3\textwidth,angle=-90]{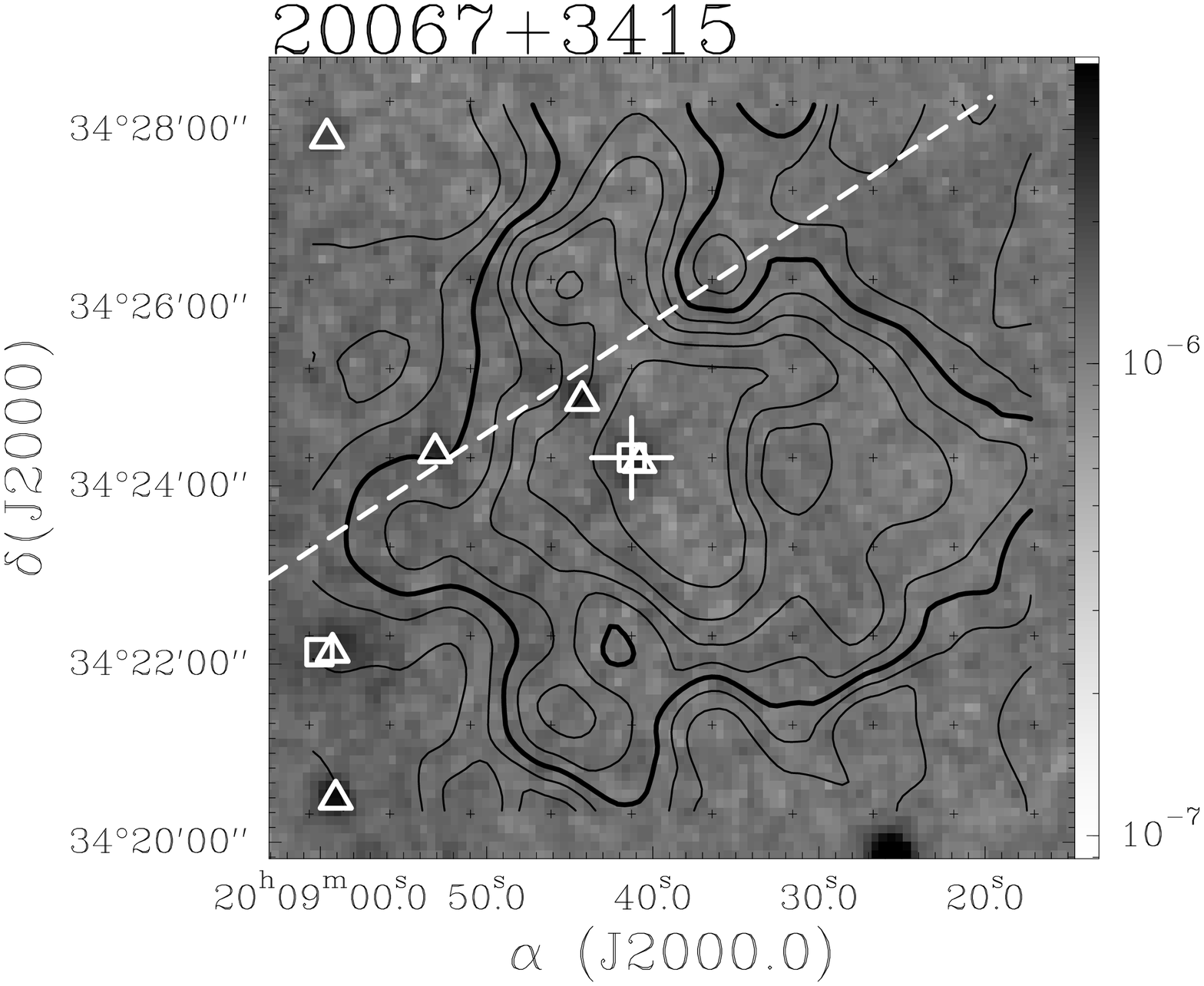}}\par
\vskip -0.01\textwidth
\subfloat{\includegraphics[width=0.235\textwidth,height=0.3\textwidth,angle=-90]{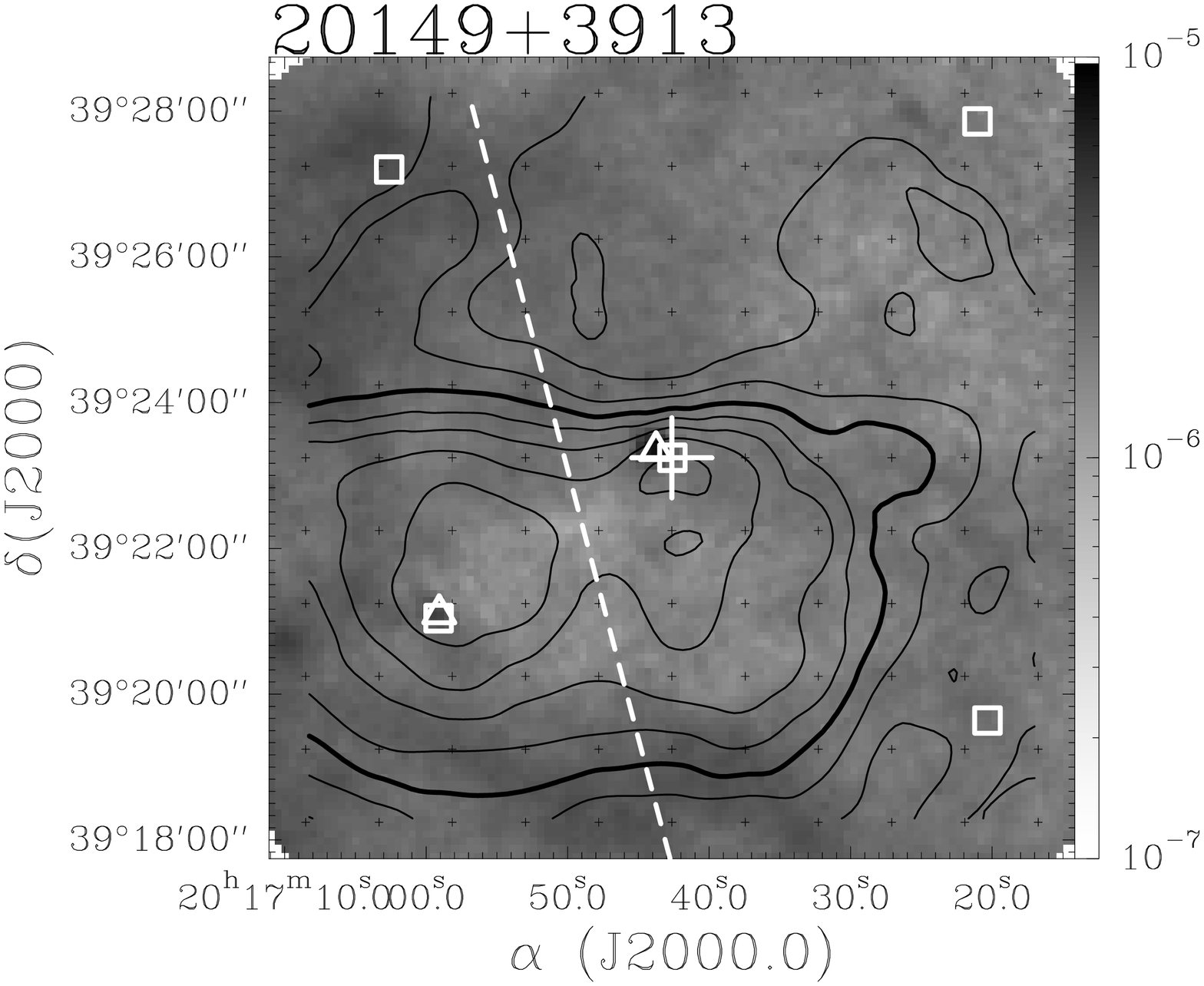}}
\subfloat{\includegraphics[width=0.235\textwidth,height=0.3\textwidth,angle=-90]{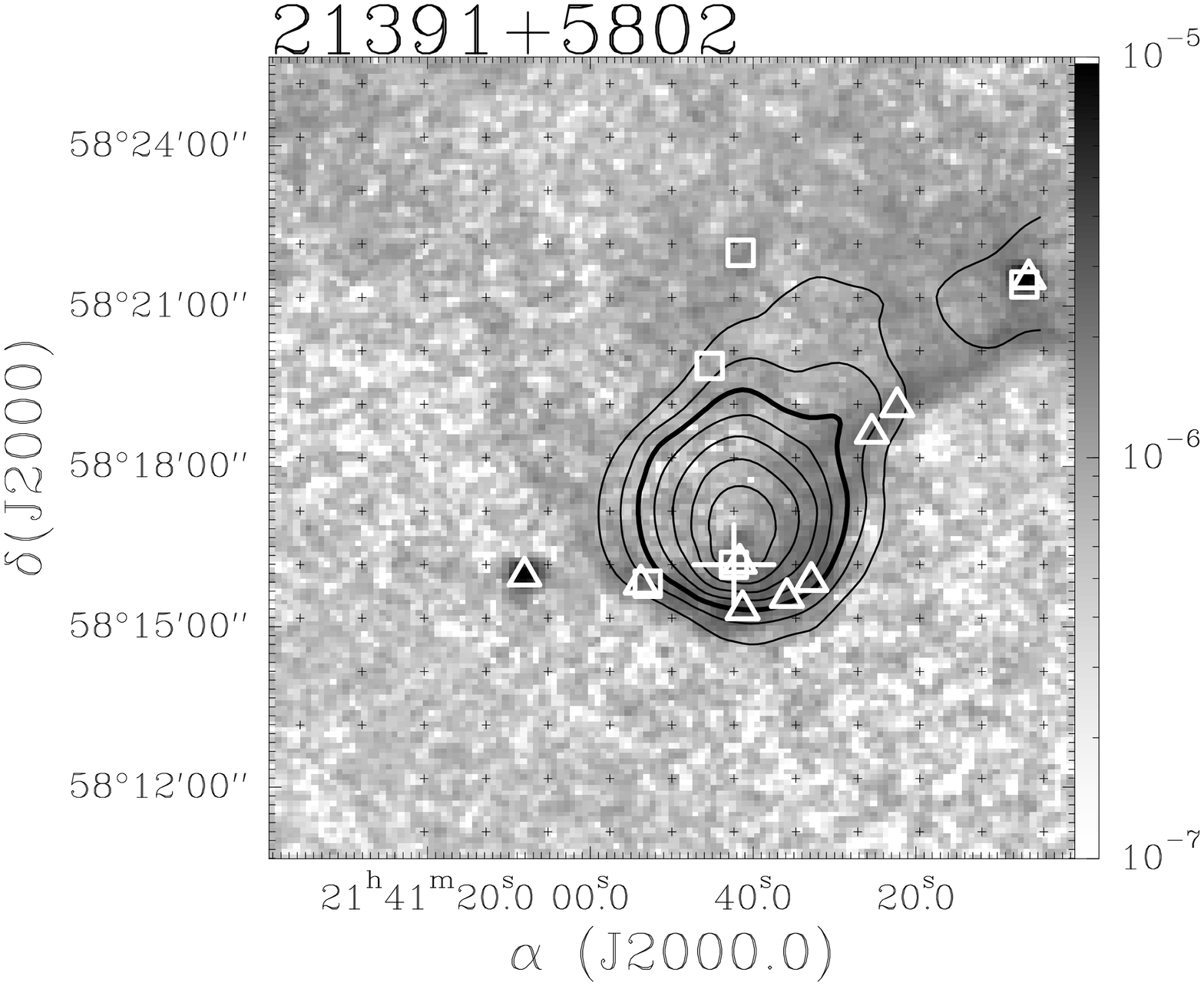}}
\subfloat{\includegraphics[width=0.235\textwidth,height=0.3\textwidth,angle=-90]{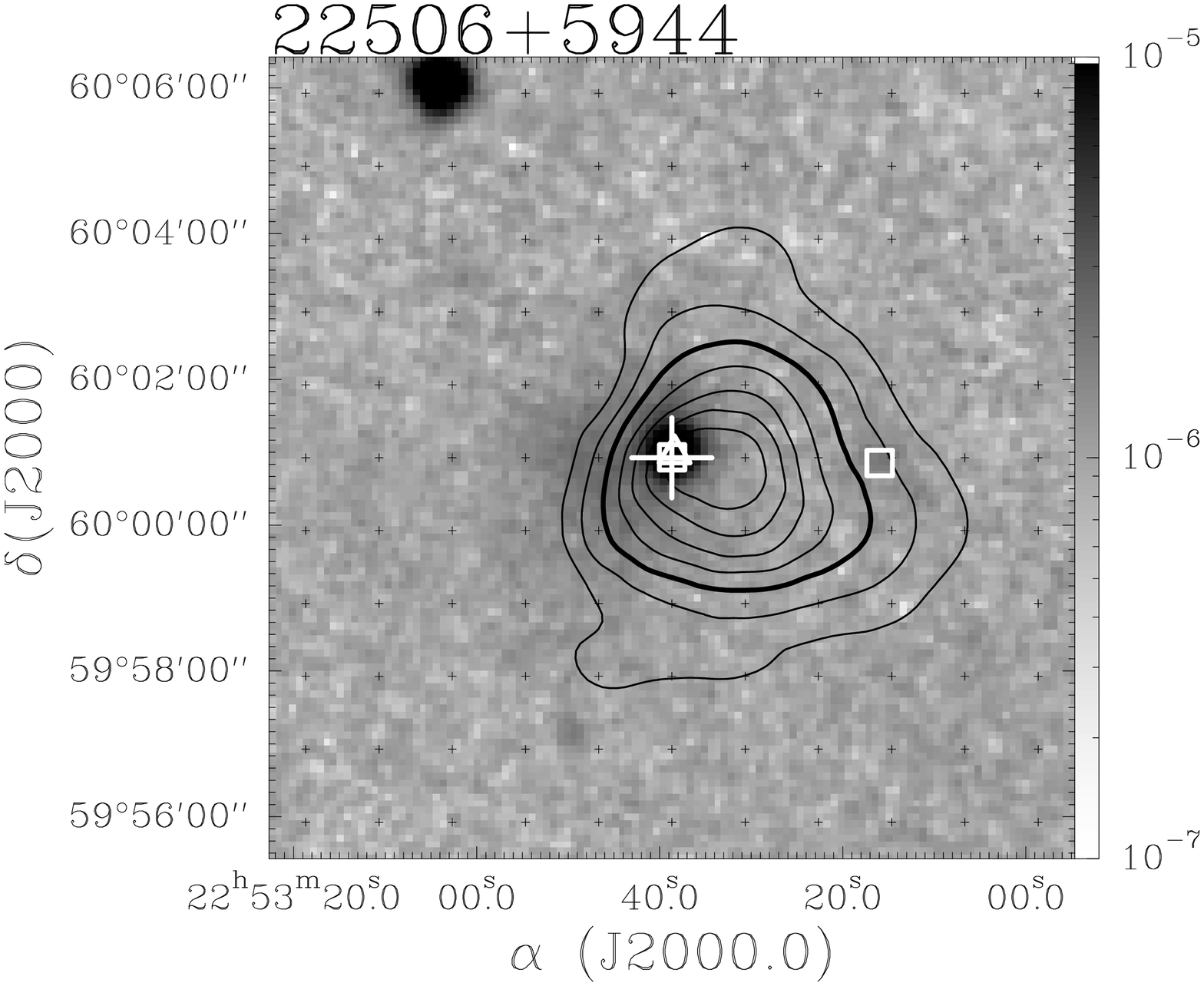}}\par
\vskip -0.003\textwidth
\subfloat{\includegraphics[width=0.235\textwidth,height=0.3\textwidth,angle=-90]{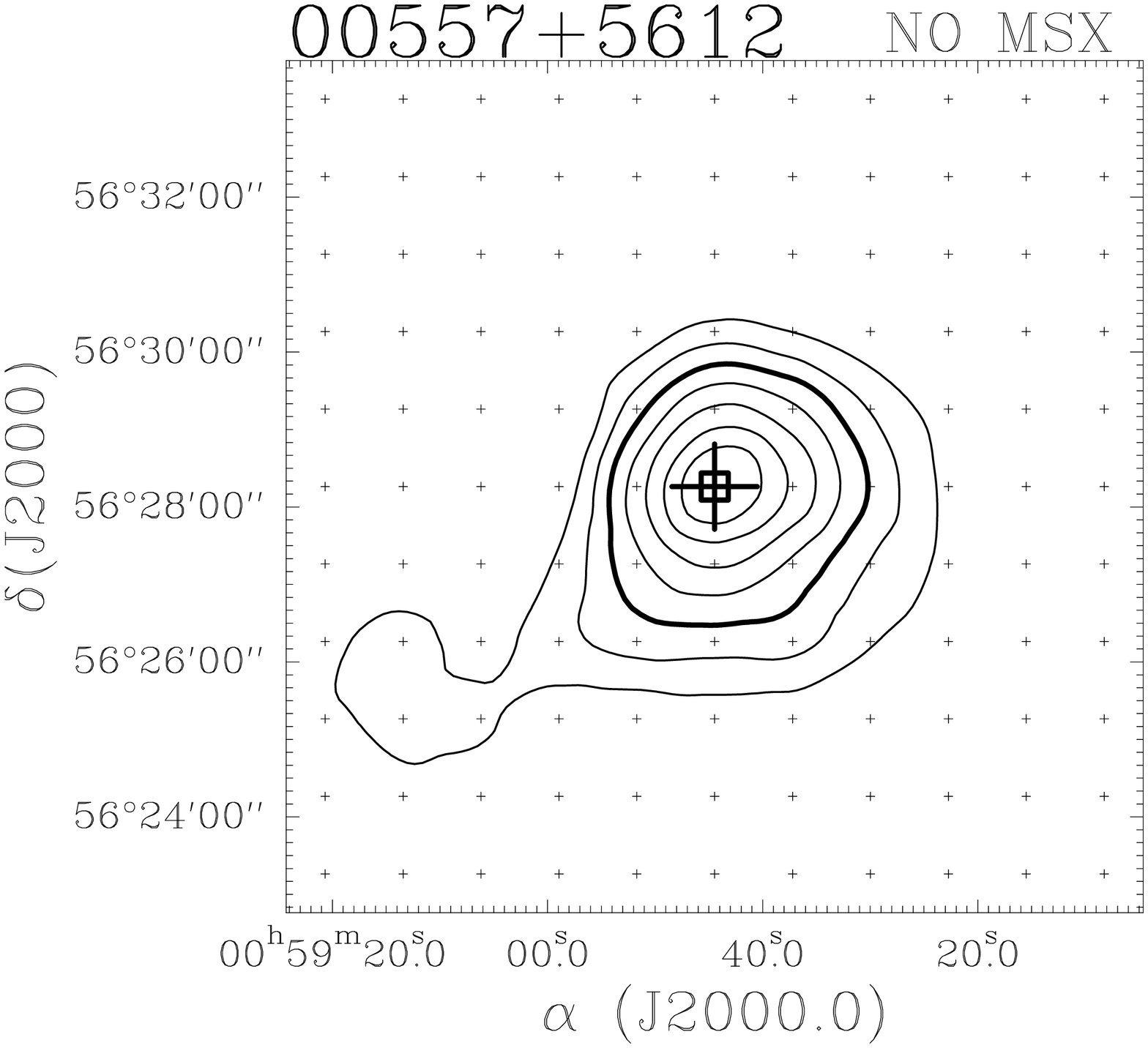}}\hspace{0.011\textwidth}
\subfloat{\includegraphics[width=0.235\textwidth,height=0.3\textwidth,angle=-90]{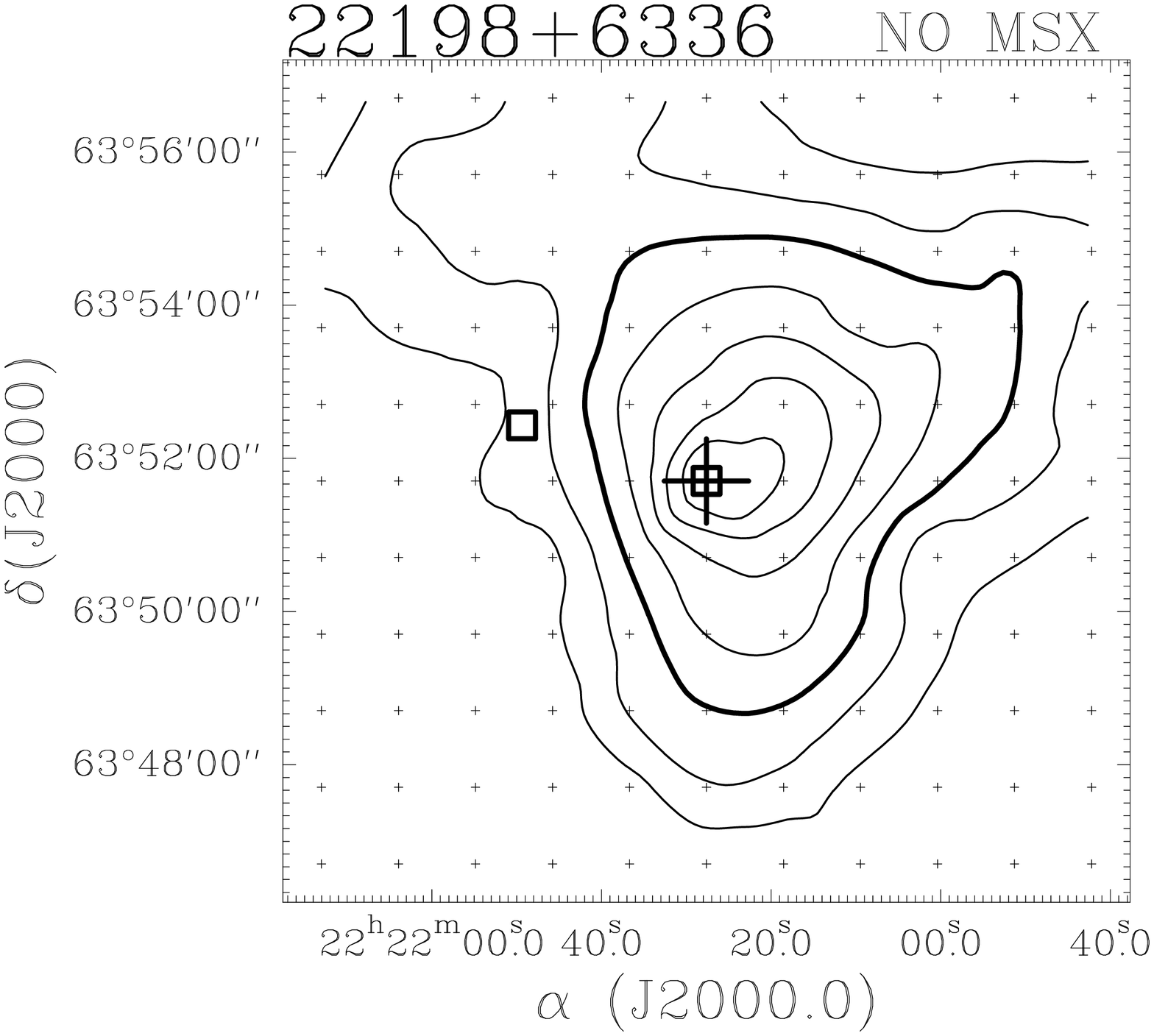}}
\caption{ $^{13}$CO~\textit{J}\,=\,2\,--\,1  integrated intensity
contours overlay the \textit{Midcourse Space Experiment}
(\textit{MSX}; \citealt{2001AJ....121.2819P}) band A (8.28\,$\mu$m)
images as background, if available. Maps are listed in order of
name, except the two without \textit{MSX} data. \textit{IRAS}
03258+3104 has no band A data and we use band C instead;
\textit{IRAS} 00557+5612 and 22198+6336 both have no \textit{MSX}
data available, as labeled on the top of their relevant sub-figures.
\textit{IRAS}  point sources are denoted by squares while
\textit{MSX}  point sources by triangles. Small crosses represent
observed points, while large crosses denote the original guide
sources (Sect.\,\ref{subsect:map}), as labeled on the top of each
map. Dash lines schematically separate resolved cores. For
integration, only intensity over three times of the standard
deviation ($3\sigma$) is considered. Contour levels begin from 30\%
to 90\% by 10\% of the peak intensity, while 50\% level is
highlighted by a solid thick contour. For \textit{MSX} images, the
grey scale wedge is shown on the right side; the unit is $W\,
\rm{m}^{-2}\,\rm{sr}^{-1}$. \label{fig:maps}}
\end{figure*}

To improve our understanding of the properties of the surveyed
sample, 14 sources were mapped in $^{13}$CO~\textit{J}\,=\,2\,--\,1
and compared with archival mid-infrared (MIR) continuum data. Mapped
sources were selected from the surveyed sample as those with only
single emission component, and they almost evenly cover longitude
$70^{\circ}<l<230^{\circ}$, avoiding low Galactic longitudes, where
$^{13}$CO lines are often affected by multiple velocity components
from the Galactic molecular ring. Using these sources as a guide,
maps were extended until at least one core was resolved. We name a
map on the basis of its guide source name, as outlined in
Fig.~\ref{fig:maps}. In four cases, one map resolved two cores,
resulting in 18 cores in total. We found that 13 cores are
associated with the original guide sources, two cores are associated
with other \textit{IRAS} sources, and three cores have no embedded
infrared source (sourceless hereinafter). A core is named after its
associated \textit{IRAS} source; for a sourceless core, it is named
after its nearest \textit{IRAS} source plus relative direction to
the core (e.g., 20067+3415NE). See Table~\ref{tab:map14} for core
properties.

The core size (Col. 2 of Table~\ref{tab:map14}) is defined as an
equivalent linear size $R = \sqrt{A/\pi}$, where $A$ is the
projected area of each cloud within the 50$\%$ contour (highlighted
in Fig.~\ref{fig:maps}). It is corrected for the effect of beam
smearing by multiplying its value by a factor
${\sqrt{\theta_{\rm{obs}}^2-\theta_{\rm{mb}}^2}}/{\theta_{\rm{obs}}}$,
where $\theta_{\rm{obs}}$ is the angular diameter of the core and
$\theta_{\rm{mb}}$ is the beamwidth. For three cores, the observed
angular diameters are comparable to the beamwidth, so that the cores
are just marginally resolved and the corresponding core sizes are
highly uncertain. In a few cases, maps were not complete to 50$\%$
of the peak intensity, and can only infer lower limits to $R$
(indicated by a symbol '$>$'). The average line width of each core
(Col. 3) is determined by combining all the spectra in the core and
then fitting a Gaussian profile to the average spectrum. In a few
cases, the average spectra show line asymmetry/absorption and need
to be fitted with two Gaussian profiles, and then the line width of
the stronger profile is given. The typical uncertainty in the
average line width is 0.04\,$\rm{km\,s}^{-1}$. Column (4) lists the
luminosity also given in Table~\ref{tab:survey98} for reference.
Peak volume densities for H$_2$, $n(\mathrm{H}_{2})$ (Col. 5), and
the LTE core masses, $M_\mathrm{LTE}$ (Col. 6), are calculated based
on both $R$ and the peak $^{13}$CO column densities determined by
interpolating the maps. For three maps (\textit{IRAS} 06067+2138,
07024$-$1102, and 21391+5802), however, no $N(^{13}\mathrm{CO})$ are
available in Table~\ref{tab:survey98} because of a lack of $^{12}$CO
data. To estimate their core properties, we assume reasonable
excitation temperatures: for \textit{IRAS} 06067 and 07024, we
assume a typical $T_\mathrm{ex}$ of 15\,K; and for \textit{IRAS}
21391, we assume that $T_\mathrm{ex}$ equals the dust temperature
(25\,K, \citealt{2002ApJ...573..246B}). Column (7) presents the
virial mass derived from the sizes and line widths following
\cite{1988ApJ...333..821M}. The ratio of virial to LTE mass $\alpha
= M_\mathrm{vir}$/$M_\mathrm{LTE}$ is listed in the last Col. (8) of
Table~\ref{tab:map14}. We exclude the marginally resolved cores when
computing averages except for the line width column.

Overall, the core mass ranges from $\sim 10^2 M_{\odot}$ to $10^4
M_{\odot}$, the linear size from 0.11\,pc to 2.41\,pc, and molecular
hydrogen density is in the range $\sim 10^3-10^4\,\rm{cm}^{-3}$. The
luminosities are once again, distributed across a wide range, from
30\,$L_{\odot}$~to 1.7$\times10^4 L_{\odot}$. Overall, the line
width $\overline{\Delta V}_{13} >\, \sim2\,\rm{km\,s}^{-1}$, and has
an average of 2.80\,$\rm{km\,s}^{-1}$, smaller than that of the
entire surveyed sample. We find an average value of 1.3 for the
ratio of virial to LTE core mass, $\alpha$. Overall the mapping
sample infers $\alpha\sim 1$, indicating that most of the cores
appear to be virialized.

Comparisons between $^{13}$CO maps and MIR images are presented in
Fig.~\ref{fig:maps}, and a detailed evolutionary identification of
individual mapped sources is presented in the online
\mbox{Appendix~A.}

\section{Discussion} \label{sect:discussion}

\subsection{The line width-luminosity relation}
\label{sect:LL criteria}

The empirical correlation of line width versus luminosity has been
found by various authors from observations in C$^{18}$O 
\citep{2001PASJ...53.1037S, 2003AJ....126..286R}, as well as in NH$_3$
\citep{1988A&A...203..367W, 1991ApJ...372L..95M, 1993A&AS...98...51H, 1994ApJ...433..117L, 1999ApJS..125..161J}. 
The line width is in general found to increase
with luminosity, for different $\lg(\Delta
V)/\lg(L_{\rm{bol}})$ slopes: 0.13-0.19 for NH$_3$
\citep{1999ApJS..125..161J} and 0.11 for C$^{18}$O
\citep{2001PASJ...53.1037S}. Given the relatively wide availability
of the archival $^{13}$CO data, it is helpful to compile an
up-to-date $^{13}$CO observed sample to investigate the
luminosity-line width relation in case of $^{13}$CO.

\begin{figure*}[htb]
 \centering
 \includegraphics[width=12cm]{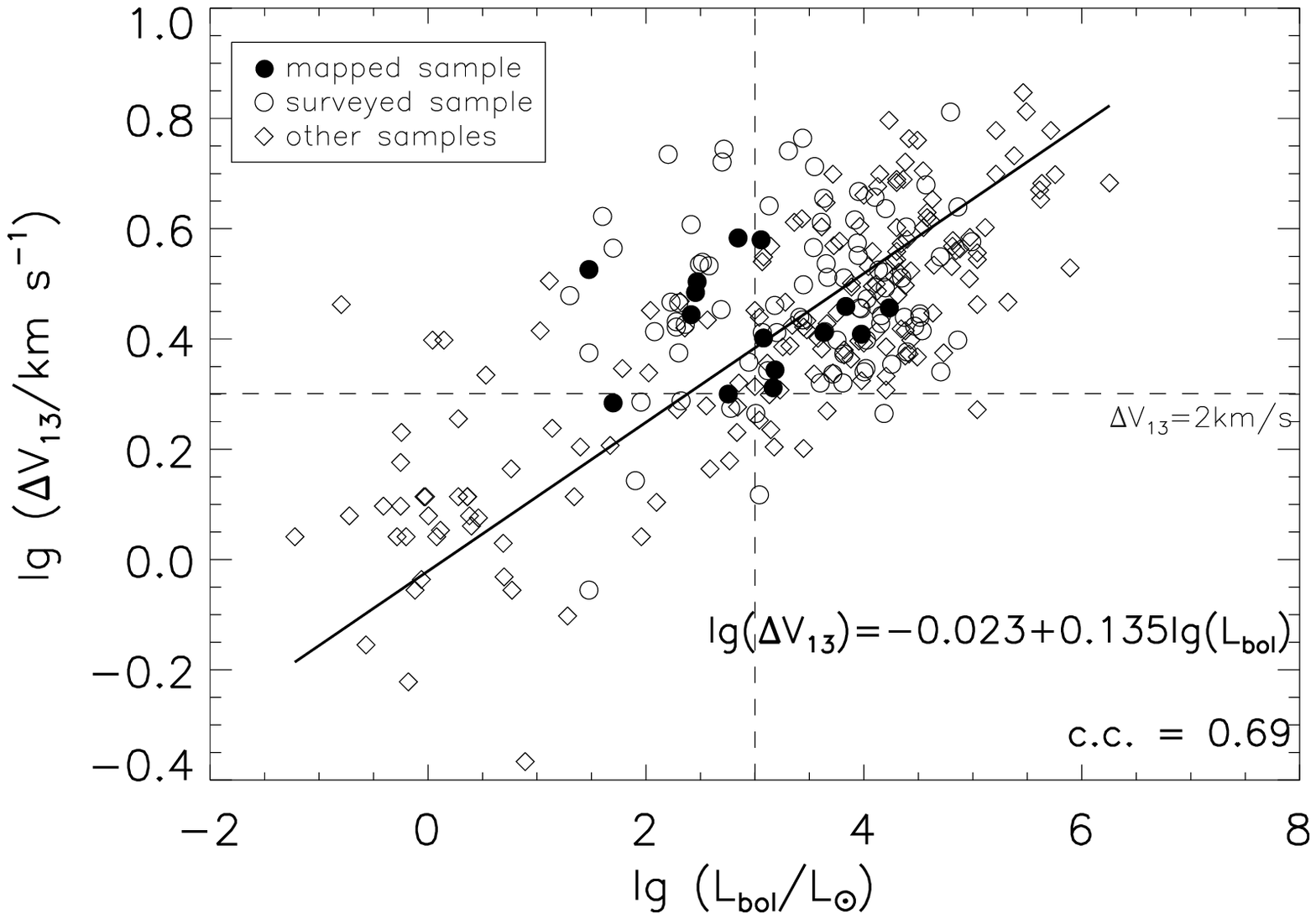}	
\caption{ Line width plotted versus bolometric luminosity of 360
$^{13}$CO observed sources: surveyed sources (open circles), mapped
sources (filled circles), and other samples adopted from the
literature (diamonds). Solid line represents a least squares fitting
to the data, and the dash lines represent the luminosity/line-width
criteria (Sect.\,\ref{sect:LL criteria}).} \label{fig:lwidth-lum}
\end{figure*}
In Fig.~\ref{fig:lwidth-lum}, we plot in logarithmic space the
line width versus luminosity from our sample and other $^{13}$CO
observed samples adopted from the literature
\citep{1986ApJ...307..337B, 1985A&A...146..375D,
1985ApJ...293..508F, 1989ApJ...347..894Y, 2001A&A...380..665W,
2004A&A...426..503W, 2003AJ....126..286R}. This sample contains 360
sources in total. One finds that the luminosity of the \textit{IRAS}
sources is well correlated with the $^{13}$CO line width, as fitted
by a power law:
\begin{displaymath}
\lg\,\bigg(\frac{\Delta V_{13}}{\rm{km\,s}^{-1}}\bigg) = (-0.023\pm
0.044)+(0.135\pm 0.012)\lg\,\bigg(\frac{L_{\rm
bol}}{L_{\odot}}\bigg),
\end{displaymath}
where the correlation coefficient $\rm{c.c.=0.69}$. This suggests
that the mass of the forming stellar objects is linked to the
dynamic status of their parent clouds. \cite{2001PASJ...53.1037S}
measured a similar correlation in the Centaurus tangential region
and suggested that the mass of the formed stars is determined by the
internal velocity dispersion of the dense cores. If the velocity
dispersion is a reliable indicator of the turbulence, this is
consistent with the idea that turbulence is different in high-mass
and low-mass cores. We note that the $^{13}$CO line widths adopted
from the literature were measured from \textit{J}\,=\,1\,--\,0
transition with smaller beams.

It is generally agreed that embedded infrared point sources of high
luminosity ($\geq$\,$10^{3}L_{\odot}$) but without associated
H\,{\sc ii} regions are good candidates to be high-mass YSOs in the
pre-UC H\,{\sc ii} phase (e.g., \citealt{2006A&A...450..607W} and
references therein). However, sufficiently young massive objects are
not necessarily bright at infrared wavelengths; some of them have no
infrared counterparts. High-mass objects at sufficiently young
evolutionary stages could be quite faint in infrared ranges either
because they are not yet mature enough to have developed infrared
emission or they are embedded very deeply in cold dust. For
instance, \citet{2005ApJ...634L..57S} identified several 1.2\,mm
emitting high-mass starless cores (HMSCs) that exhibit absorption or
no emission at the MIR wavelengths; a centimeter-emitting UC H\,{\sc
ii} region was also found without an infrared counterpart
\citep{2008A&A...477..267F}. In our mapped sample, faint
($<10^{3}L_{\odot}$) sources associated with very massive
($>10^{3}M_{\odot}$) cores (20149, 20151) do exist. Although we
cannot rule out the possibility that these two sources could evolve
to only low-mass stars, it is very likely that the clouds will
eventually fragment to form high-mass stars, given the large amount
of gas therein.

Hereafter, for clarity, our \emph{luminosity criterion} refers to
bolometric luminosity $L_{\rm bol} \geq 10^{3}\, L_{\odot}$, and our
\emph{line-width criterion} refers to line width $\Delta V
$($^{13}$CO~\textit{J}\,=\,2\,--\,1)$\,>2$\,$\rm{km\,s}^{-1}$.
According to the $\Delta V - L$ relation above, luminosity $L_{\rm
bol}=10^{3}\,L_{\odot}$ corresponds to $\Delta V_{13} =
2.42^{+0.49}_{-0.41}\,\rm{km\,s}^{-1}$. Because high-mass stars are
far more luminous than their low-mass counterparts, luminous \textit{IRAS} sources
($\geq10^{3}\,L_{\odot}$) are likely to be high-mass stellar
objects. Therefore, we tentatively (not exclusively) suggest the
lower limit, $\Delta V_{13} =2\,\rm{km\,s}^{-1}$, as a
characteristic value for the line-width criterion, analogous to the
widely used luminosity criterion. Objects with line width larger
than this characteristic value are probably high-mass objects. The
line-width criterion includes 94.5\% sources that also satisfy the
luminosity criterion (Fig.~\ref{fig:lwidth-lum}), which implies
that line width may be a key parameter in measuring the masses of
the forming stellar objects in the cores, at least in our sample. We
note that this criterion (2\,$\rm{km\,s}^{-1}$) is larger than the
typical line width of low mass cores (1.3\,$\rm{km\,s}^{-1}$,
\citealt{1983ApJ...264..517M}), and smaller than the average line
width of high-mass cores (3.5\,$\rm{km\,s}^{-1}$,
\citealt{2001A&A...380..665W}).

Applying the luminosity/line-width criteria to our sample, there are
68 sources satisfying luminosity criterion, 65 (95.6\%) of which
also satisfy line-width criterion. We suggest that the 65 sources
are candidate high-mass star formation regions in a pre-UC H\,{\sc
ii} phase. For the remaining 30 less luminous sources, 23 of them
satisfy the line-width criterion but not the luminosity criterion.
We suggest that the 23 sources are high-mass YSO candidates earlier
than pre-UC H\,{\sc ii} phase.

\subsection{Core masses and line widths} \label{sect:map discussion}

The core masses provide a direct test of our line-width criterion.
Figure~\ref{fig:lwidth-lum} includes 15 cores with luminosities
listed in Table~\ref{tab:map14}. We exclude sourceless cores and
marginally resolved cores in the discussion in this section, because
the former's luminosity cannot be determined and latter's estimated
size and mass are highly uncertain. The remaining 14
luminosity-available cores can be divided into two groups: Group I,
which do not satisfy the luminosity criterion, including 03414,
21391, 03260, 20149, 06067, 20151; and group II, satisfying the
luminosity criterion, including 00557, 22198, 03101, 06103, 00117,
05168, 22506, 20067 (in order of increasing line width). All group
II cores also satisfy the line-width criterion, and they are located
in the upper right panel of Fig.~\ref{fig:lwidth-lum}. They are
very massive, their estimated masses being higher than several
$10^2$$M_{\odot}$, except core 00117 ($1.8\times10^2 M_{\odot}$). On
the other hand, group I cores mostly satisfy the line-width
criterion. They are relatively less massive than group II cores,
their masses being typically $\sim 10^2 M_{\odot}$ or more, with two very
massive cores 20149 and 20151 ($>$\,$10^3 M_{\odot}$). The only case
that does not satisfy the line-width criterion, core 03414, has the
lowest mass in group I. We note that all group I cores are still
significantly more massive than the low-mass cores
\citep{1983ApJ...264..517M}.

The high-mass nature of group I cores confirms again (in addition to
the previously mentioned cores 20149, 20151) that the luminosity
criterion cannot be applied to some young sources. On the other
hand, the line-width criterion is applicable to our sample. While in
terms of inferring mass, line width may not be as direct an
indicator as luminosity, it is helpful when luminosity is
unavailable or is affected by large uncertainty (e.g., due to
distance ambiguity, flux upper limit), which is often the case. In
addition, line width can be measured observationally more easily and
more accurately than luminosity.

\begin{figure}[!t]
 \centering
 \includegraphics[width=0.5\textwidth]{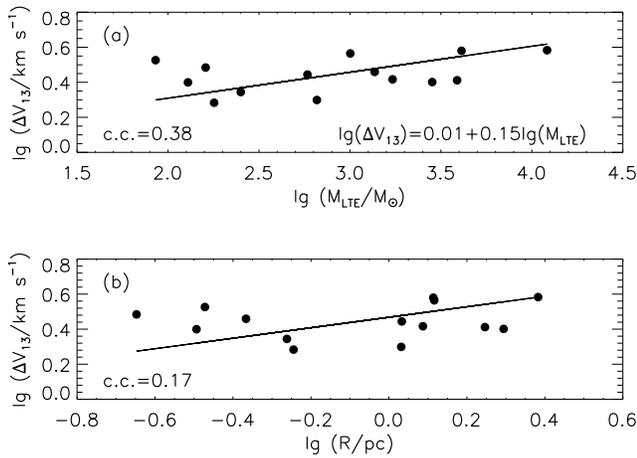}
\caption{ Relations between $^{13}$CO line width $\Delta V_{13}$ and
other physical parameters of molecular cores: (a) LTE mass
$M_{\mathrm{LTE}}$, and (b) size $R$. In both cases, the filed
circles represent data and solid lines are linear fitting results to
the data. The fitting function and correlation coefficient are
labeled in each panel.} \label{fig:corrs}
\end{figure}

In Fig.~\ref{fig:corrs} (a), we plot $\lg (\Delta V_{13})$ versus
$\lg M_{\mathrm{LTE}}$. A weak correlation is evident in the data,
with a correlation coefficient of 0.38. This may indicate that, for
molecular clouds with associated YSOs, the $^{13}$CO line width at
some degree is related to the cloud mass, and massive cores tend to
have larger line widths. This weak correlation, together with the
strong $\Delta V - L$ correlation, indicates that massive stars are
more likely to form in massive molecular cores. The core mass and
associated \textit{IRAS} luminosity in our mapped sample are indeed
well correlated ($\rm{c.c.}=0.76$). However, this correlation may
need to be corrected for distance effects (0.3--6.08\,kpc for mapped
sample) because both mass and luminosity are proportional to $D^2$.
Nevertheless, strong mass-luminosity correlations were reported in
other regions or samples that have far smaller distance differences
than our sample \citep{1996ApJ...466..282D, 2001PASJ...53.1037S,
2003AJ....126..286R}. Figure.~\ref{fig:corrs} (b) plots line width
$\Delta V_{13}$ and core size $R$ in logarithm. With an average
virial mass of $1.1\times10^3\,M_{\odot}$, the cores exhibit no
correlation between size and line width. This indicates that the
Larson law is invalid for our mapped sample, consistent with the
results of previous works (\citealt{1997ApJ...476..730P}, $\langle
M_{\mathrm{vir}} \rangle > 3.8\times10^3\,M_{\odot}$;
\citealt{2008MNRAS.391..869G}, $\langle M_{\mathrm{vir}} \rangle =
5.6\times10^3\,M_{\odot}$). The correlations in Fig.~\ref{fig:corrs}
are unaffected by distance, because the line width and distance of
the plotted cores are not correlated ($\rm{c.c.}=0.04$).

We conclude that, based on the currently available sample, YSOs with
higher bolometric luminosity ($>10^3\,L_{\odot}$), tend to be
associated with more massive molecular cloud structures, which are
usually more turbulent, and have a large $^{13}$CO line width,
$\Delta V_{13}>2\,\rm{km\,s}^{-1}$. It is important to note that,
the characteristic value ($2\,\rm{km\,s}^{-1}$) may not be
universal, and can vary from region to region and/or from line to
line. Further mapping of more clouds, as well as higher angular
resolution data if available, are required to examine the line-width
criterion proposed here.

\section{Summary} \label{sect:summary}

We have carried out a \mbox{$^{13}$CO~\textit{J}\,=\,2\,--\,1}
survey of 135 \textit{IRAS} sources selected as potential YSOs
earlier than UC H\,{\sc ii} regions. Our main findings are
summarized as follows:

1. Ninety-eight sources have good enough emission profile for
analysis; some of them show asymmetric line profiles of
$^{13}$CO~\textit{J}\,=\,2\,--\,1. The line width is
3.09\,$\rm{km\,s}^{-1}$ and excitation temperature is 9.7\,K,  on average. 
The H$_2$ column densities are
$(1.7-40.8)\,\times$$10^{21}\,\mathrm{cm}^{-2}$. Sixty-five sources
are suggested to be candidate precursors of UC H\,{\sc ii} regions.

2. Fourteen sources were mapped and resolved as eighteen cores,
which have been identified with eight pre-UC H\,{\sc ii} regions and
one UC H\,{\sc ii} region, two high-mass cores earlier than pre-UC
H\,{\sc ii} phase, four possible star forming clusters, and three
sourceless cores.

3. For molecular clouds with known associated YSOs and measured
$L_{\rm bol}$, $^{13}$CO line width $\Delta V_{13}$ of the clouds is
correlated with the bolometric luminosity of the YSOs. Based on the
current $^{13}$CO observed sample (360 sources in total), this
correlation can be fitted as a power law, $\lg\,(\Delta
V_{13}/\,\rm{km\,s}^{-1}) = (-0.023\pm 0.044)+(0.135\pm
0.012)\lg\,(L_{\rm bol}/L_{\odot})$.

4. Luminous ($>10^3\,L_{\odot}$) YSOs tend to be produced in more massive and more turbulent
($\Delta V_{13}>2\,\rm{km\,s}^{-1}$) molecular cloud structures.

5. High-mass stars are more likely to form in massive molecular
clouds.

\begin{acknowledgements}
We are grateful to Q.~Zhang, X.~Guan, R.~Xue and L.~Zhu for valuable
discussion. We thank H.~Du and F.~Virgili for their help on the
manuscript. We also thank the anonymous referee whose comments and
suggestions helped to improve the content and the clarity of this
paper. This research is supported by Grants 10733030 and 10873019 of
NSFC. It made use of data products from the Infrared Astronomical
Satellite (\textit{IRAS}) and the Midcourse Space Experiment
(\textit{MSX}) retrieved from the NASA/IPAC Infrared Science
Archive, which is operated by the JPL/Caltech under a contract with
NASA.
\end{acknowledgements}

%\bibliographystyle{aa}
%\bibliography{astro}

%\appendix
\Online
\begin{appendix}
\section{Individual Analyses} \label{sect:map analysis}

In Fig.~\ref{fig:maps}, we compare the integrated intensity of
$^{13}$CO emissions to \textit{IRAS} (point sources) and \textit{MSX} 
(point sources and images) data. We use the most sensitive band
(band A, centered on 8.28\,$\mu$m) images of \textit{MSX} when
available. We note that maps in Fig.~\ref{fig:maps} are labeled by
the original guide sources (Sect.\,\ref{subsect:map}). \textit{IRAS}
03258+3104 has no band A data and we use band C (centered on
12.13\,$\mu$m) instead. \textit{IRAS} 00557+5612 and 22198+6336 both
have no \textit{MSX}
data. Individual analyses of each map are presented as follows.\\

\indent{\textit{\object{IRAS 00117+6412}:}} $^{13}$CO emission peak coincides
well with \textit{IRAS} and \textit{MSX} point sources, and there
are also strong counterparts in all four \textit{MSX}  bands. This
distinctive $^{13}$CO core is massive ($1.8\times10^2 M_{\odot}$),
in agreement with the conclusion deduced from the
luminosity/line-width criteria (Sect.\,\ref{sect:LL criteria}). We
therefore suggest that it is a pre-UC H\,{\sc ii} region. Strong 22
GHz water maser \citep{1993A&AS...98..589W} and outflow activity
\citep{2005ApJ...625..864Z, 2003AcASn..44S.103Z}
have been detected within this area, providing evidence of active star formation.\\

\indent{\textit{\object{IRAS 00557+5612}:}} \textit{MSX}  data are
unavailable close to this region, but the \textit{IRAS}  source
matches well to the $^{13}$CO core peak. A core as massive as
$6.6\times10^2 M_{\odot}$ agrees with the conclusion deduced from
the luminosity/line-width criteria. We therefore suggest that it is
a pre-UC H\,{\sc ii}  region. A velocity gradient of
0.36\,$\rm{km\,s}^{-1}$\,$\rm{pc}^{-1}$ from northeast to southwest
is inferred, yielding a rotating angular velocity of
$1.16\times10^{-14}\,\rm{s}^{-1}$. According to
$^{13}$CO~\textit{J}\,=\,1\,--\,0  and HCO$^{+}$ mapping by
\cite{2007ChJAA...7..331Z}, two subcores
exist within this core.\\

\indent{\textit{\object{IRAS 03101+5821}:}} $^{13}$CO emission coincides with
infrared point sources and image as well. A core more massive than
$5.8\times10^2 M_{\odot}$ agrees with the conclusion drawn from the
luminosity/line-width criteria. We therefore suggest that it is a
pre-UC H\,{\sc ii} region. A 22 GHz water maser has been detected
within this area \citep{1993A&AS...98..589W}.\\

\indent{\textit{\object{IRAS 03258+3104}:}} a \textit{MSX} band A image is
unavailable for this region and we use band C instead. $^{13}$CO
emission within this area is more diffuse than that in former
sources. At least two cores (03260+3111 and 03260+3111NE) are
resolved within an area of 0.34\,pc. The larger core coincides with
\textit{IRAS} 03260+3111, which does not satisfy our color selection
criteria and was suggested as an UC H\,{\sc ii} region by
\cite{1990A&AS...83..119C}. Taking both their line width and
luminosity into account, we suggest that core 03260+3111 is a
high-mass object in UC H\,{\sc ii} phase, while 03260+3111NE is a
sourceless core. The original guide source of the map, \textit{IRAS}
03258+3104, is not associated with any resolved $^{13}$CO core. It
has been suggested to be a Class 0 object driving a low-mass bipolar
CO outflow \citep{2000A&A...361..671K}.\\

\indent{\textit{\object{IRAS 03414+3200}:}} $^{13}$CO is quite diffuse
across the entire area of 0.34\,pc, so that no distinctive $^{13}$CO
core is found. However, several infrared point sources are evident
close to the 90\% contour within one beam, superimposed on a steep
density gradient. Although its mass ($>$\,85\,$M_{\odot}$) and line
width 1.92\,$\rm{km\,s}^{-1}$) are relatively low
in all the mapped sources, it appears to be a star forming cluster.\\

\indent{\textit{\object{IRAS 05168+3634}:}} at least two cores (05168+3634
and 05168+3634SW) are present within 3.01\,pc. The dominant
northeastern core appears to be associated with the infrared
sources. The total core mass higher than $1.5\times10^4 M_{\odot}$
agrees with the conclusion deduced from the luminosity/line-width
criteria. We suggest that the dominant core is a high-mass star
forming region in pre-UC H\,{\sc ii} phase, and the SW core is a
sourceless core. This is consistent with results from
\citet{1996A&A...308..573M}. 
Strong outflow activity was identified \citep{1994A&AS..103..541B, 2005ApJ...625..864Z}
at the position of the NE core, and the outflow driving source appears deviated to the infrared
source \textit{IRAS} 05168+3634. 
A 22\,GHz water maser was detected by \cite{1991A&A...246..249P} in this region.\\

\indent{\textit{\object{IRAS 06067+2138}:}} $^{13}$CO core coincides with the
\textit{IRAS}  point source but is without \textit{MSX} counterpart.
The core mass $M_\mathrm{LTE}$ ($2.5\times10^2 M_{\odot}$) is only
one third of its virial mass $M_\mathrm{vir}$ ($7.8\times10^2
M_{\odot}$), indicating that the core is not yet gravitationally bound. Taking
its large line width (3.36\,$\rm{km\,s}^{-1}$) into account, we
suggest that it is a high-mass object earlier than pre-UC H\,{\sc ii} phase.\\

\indent{\textit{\object{IRAS 06103+1523}:}} $^{13}$CO emission coincides with
both infrared point sources and image. A core as massive as
$2.8\times10^3 M_{\odot}$ agrees with the conclusion deduced from
the luminosity/line-width criteria. \textit{IRAS} 06103+1523 is
found to be two point sources by \textit{MSX}  data, implying that a
fine structure may exist there. A denser molecular tracer (e.g.,
N$_2$H$^+$ or HCO$^+$) and a higher resolution (several arcsec) are
needed to study these fine structures.
We therefore suggest that it is a high-mass star forming cluster.\\

\indent{\textit{\object{IRAS 07024-1102}:}} $^{13}$CO map is incomplete
but a core is clearly evident. The core coincides with the
\textit{IRAS} point source but does not have a \textit{MSX}
counterpart. Although its luminosity (570\,$L_{\odot}$) is a little
lower than the luminosity criterion, it does generate a line width
(1.99\,$\rm{km\,s}^{-1}$) very close to the line-width criterion. We
therefore suggest that it is a high-mass object earlier than pre-UC H\,{\sc ii} phase.\\

\indent{\textit{\object{IRAS 20067+3415}:}} the $^{13}$CO gas distribution is
quite complex in this region. At least two cores (20067+3415 and
20067+3415NE) are revealed. The dominant southwestern core
coincides with infrared point sources, while the northeastern core
is a sourceless core. Several sub-structures are revealed within an
area of 2.21\,pc with total mass of $4.9\times10^3 M_{\odot}$. Thus,
we suggest that core 20067+3415 is a high-mass star forming cluster.
The MIR emission and luminosity are relatively weak compared to
those of other pre-UC H\,{\sc ii} regions, indicating a very early evolutionary stage.\\

\indent{\textit{\object{IRAS 20149+3913}:}} $^{13}$CO emission reveals two
cores (20149+3913 and 20151+3911); both also coincide with infrared
point sources and image. They do not satisfy the luminosity
criterion but have large line widths, consistent with their high
masses. Both sources are located in the \mbox{Cygnus X} molecular
cloud complex and were mapped in 1.2\,mm continuum
\citep{2007A&A...476.1243M}, yielding masses 8\,$M_{\odot}$ and
23\,$M_{\odot}$, respectively (see their Table 1 and Fig. 13).
Here we suggest that both cores are pre-UC H\,{\sc ii} regions.\\

\indent{\textit{\object{IRAS 21391+5802}:}} $^{13}$CO emission detects a
distinctive core and a belt of gas distributed along the southeast
to northwest direction, which is coincident with the MIR background
and several point sources. The core mass $M_\mathrm{LTE}$
($1.3\times10^2 M_{\odot}$) is roughly half of its virial mass
$M_\mathrm{vir}$ ($3.1\times10^2 M_{\odot}$), indicating that the core is not yet 
gravitationally bound, responsible for a large line width
(2.78\,$\rm{km\,s}^{-1}$). Taking its relatively small size
(0.32\,pc) into account, we suggest that it is a star forming
cluster where high-mass stars could eventually form. A 22\,GHz water
maser and outflow were identified in this region \citep{1991A&A...246..249P, 2005ApJ...625..864Z}.\\

\indent{\textit{\object{IRAS 22198+6336}:}} \textit{MSX} data are unavailable
near this region, but the \textit{IRAS} source matches the $^{13}$CO
core peak well. A core as massive as $1.7\times10^3 M_{\odot}$
agrees with the conclusion deduced from the luminosity/line-width
criteria. A 22\,GHz water maser and outflow were identified in this
region \citep{1991A&A...246..249P, 2005ApJ...625..864Z}. We
therefore suggest that it is a pre-UC H\,{\sc ii}  region.\\

\indent{\textit{\object{IRAS 22506+5944}:}} $^{13}$CO core coincides well
with a luminous \textit{IRAS} point source and a bright \textit{MSX}
counterpart. A core as massive as $1.0\times10^3 M_{\odot}$ agrees
with the conclusion deduced from the luminosity/line-width criteria.
Thus, we suggest that it is a pre-UC H\,{\sc ii} region. Outflow
activity \citep{2005AJ....129..330W,2005ApJ...625..864Z} and a
22\,GHz water maser \citep{1991A&A...246..249P} were identified,
indicating that active star formation process is underway.\\

In summary, $^{13}$CO~\textit{J}\,=\,2\,--\,1  mapping reveals at
least 18 massive cores from 14 maps. By means of individual
analyses, we identify eight pre-UC H\,{\sc ii} regions and one UC
H\,{\sc ii} region, two high-mass cores earlier than pre-UC
H\,{\sc ii} phase, four possible star forming clusters, and three
sourceless cores.

\end{appendix}

\clearpage \onecolumn

%--------------------Online table landscape-----------------
\onllongtabL{3}{
\begin{landscape}
\footnotesize
\setcounter{table}{0}
\begin{longtable}{lllll rrccc rrccc}
\caption{\label{tab:survey98}Observed and Derived Parameters
of Surveyed Sources}\\
\hline\hline\\
 {Source} &\multicolumn{6}{c}{Observed Parameters}  &  &\multicolumn{7}{c}{Derived Parameters}\\
\cline{2-7}\cline{9-15}\\
 {}                       & {$\alpha$(J2000)}          & {$\delta$(J2000)}      & {$T_\mathrm{mb12}$}     & {$T_\mathrm{mb13}$}       & {$V_\mathrm{LSR13}$}   & {$\Delta V_{13}$}    & & {$D$}    & {$L_\mathrm{bol}$}     & {$T_\mathrm{ex}$}  & {$\tau_{13}$}   & {$N(^{13}\mathrm{CO})$}     & {$N(\mathrm{H}_{2})$}   & {Ref.}\\
 {\textit{IRAS} }                  & {(h m s)}         & {($^{\circ}\,^{\prime}\,^{\prime\prime}$)}     & {(K)}    & {(K)}      & {($\rm{km\,s}^{-1}$)}     & {($\rm{km\,s}^{-1}$)} & & {(kpc)}   & {(10$^{3}\,L_{\odot}$)}    & {(K)}    & {}  & {($10^{15}\,\mathrm{cm}^{-2}$)}    & {($10^{21}\,\mathrm{cm}^{-2}$)}  & {}\\
 {(1)}                    & {(2)}           & {(3)}                                      & {(4)}               & {(5)}              & {(6)}          & {(7)}            & & {(8)}        & {(9)}                & {(10)}       & {(11)}           & {(12)}                   & {(13)}  & {(14)}\\
\hline\\
\endfirsthead
\caption{continued.}\\
\hline\hline\\
 {(1)}                    & {(2)}           & {(3)}                                      & {(4)}               & {(5)}              & {(6)}          & {(7)}            & & {(8)}        & {(9)}                & {(10)}       & {(11)}           & {(12)}                   & {(13)}  & {(14)}\\
\hline\\
\endhead
%\hline
\endfoot

00117+6412$^{m}$     &00 14 27.7       &+64 28 46            & 5.72(0.12)          & 2.55(0.05)          & -36.06(0.01)    &2.77(0.03)        &  &  3.6         &\hskip -0.3truecm  4.29              &\hskip -0.3truecm  10.50       &0.57       &\hskip -0.3truecm  5.53        &\hskip -0.5truecm  7.87       &\hskip -0.5truecm 1N,2,3N,4N,6,7,8     \\
00412+6638           &00 44 15.2       &+66 54 41            & 2.64(0.07)          & 0.81(0.14)          & -68.00(0.22)    &3.32(0.22)              &  & 7.35         &\hskip -0.3truecm 15.25              &\hskip -0.3truecm   6.96       &0.35       &\hskip -0.3truecm  3.12        &\hskip -0.5truecm  4.44       &\hskip -0.5truecm 2N,4     \\
00468+6527           &00 49 55.8       &+65 43 39            & 2.31(0.08)          & 1.17(0.07)          & -63.67(0.03)    &2.60(0.08)        &  & 6.74         &\hskip -0.3truecm 34.30              &\hskip -0.3truecm   6.55       &0.67       &\hskip -0.3truecm  4.51        &\hskip -0.5truecm  6.42       &\hskip -0.5truecm 2,4     \\
00557+5612$^{m}$     &00 58 44.4       &+56 28 16            & 8.37(0.05)          & 3.72(0.05)          & -29.96(0.01)    &2.17(0.01)        &  & 2.94         &\hskip -0.3truecm  1.47$^{u}$        &\hskip -0.3truecm  13.34       &0.57       &\hskip -0.3truecm  5.84        &\hskip -0.5truecm  8.32       &\hskip -0.5truecm 11      \\
01134+6429           &01 16 48.1       &+64 45 37            & 5.03(0.06)$^{s}$    & 1.29(0.06)          & -54.36(0.02)    &2.20(0.06)        &  &    3$^a$     &\hskip -0.3truecm  1.31              &\hskip -0.3truecm   9.73       &0.29       &\hskip -0.3truecm  2.18        &\hskip -0.5truecm  3.10       &\hskip -0.5truecm 2,4     \\
\\
02413+6037           &02 45 12.5       &+60 49 44            & 2.52(0.06)$^{s}$    & 0.93(0.04)          & -61.56(0.02)    &2.22(0.06)        &  & 7.47         &\hskip -0.3truecm 10.45$^{u}$        &\hskip -0.3truecm   6.81       &0.44       &\hskip -0.3truecm  2.58        &\hskip -0.5truecm  3.68       &\hskip -0.5truecm      \\
02485+6902           &02 53 07.2       &+69 14 36            & 4.11(0.04)$^{s}$    & 1.70(0.05)$^{s}$    & -10.52(0.01)    &1.94(0.03)        &  & 1.09         &\hskip -0.3truecm  0.21              &\hskip -0.3truecm   8.70       &0.51       &\hskip -0.3truecm  3.11        &\hskip -0.5truecm  4.43       &\hskip -0.5truecm 2N     \\
02541+6208           &02 58 13.7       &+62 20 31            & 2.28(0.06)          & 0.59(0.03)          & -51.68(0.22)    &2.09(0.22)        &  & 5.88         &\hskip -0.3truecm  6.43$^{u}$        &\hskip -0.3truecm   6.51       &0.29       &\hskip -0.3truecm  1.55        &\hskip -0.5truecm  2.20       &\hskip -0.5truecm 2,4     \\
03101+5821$^{m}$     &03 14 04.7       &+58 33 08            & 4.84(0.05)          & 2.37(0.04)          & -38.27(0.01)    &2.38(0.02)        &  &  4.2         &\hskip -0.3truecm  1.20              &\hskip -0.3truecm   9.52       &0.65       &\hskip -0.3truecm  5.03        &\hskip -0.5truecm  7.16       &\hskip -0.5truecm 2,4N,6     \\
03258+3104$^{m}$     &03 28 56.8       &+31 14 44            &13.57(0.11)          & 6.78(0.05)          &   7.56(0.01)          &3.27(0.01)        &  & 0.22$^a$     &\hskip -0.3truecm  0.03$^{u}$        &\hskip -0.3truecm  18.76       &0.68       &\hskip -0.3truecm 15.29        &\hskip -0.5truecm 21.78       &\hskip -0.5truecm 1N     \\
\\
03414+3200$^{m}$     &03 44 36.4       &+32 09 25            &17.20(0.10)          & 6.08(0.07)          &   8.40(0.01)    &1.71(0.01)        &  &  0.3$^b$     &\hskip -0.3truecm  0.05$^{u}$        &\hskip -0.3truecm  22.48       &0.43       &\hskip -0.3truecm  7.44        &\hskip -0.5truecm 10.59       &\hskip -0.5truecm 1N,2N     \\
04365+4717           &04 40 16.8       &+47 23 04            & 4.06(0.05)$^{s}$    & 1.65(0.08)          & -34.75(0.03)    &2.59(0.07)        &  &  6.9$^b$     &\hskip -0.3truecm  4.37              &\hskip -0.3truecm   8.64       &0.50       &\hskip -0.3truecm  4.04        &\hskip -0.5truecm  5.75       &\hskip -0.5truecm 2N,4     \\
05155+0707           &05 18 17.1       &+07 11 01            &  ...                & 3.34(0.06)          &  -1.45(0.01)    &2.37(0.02)        &  & 0.46$^a$     &\hskip -0.3truecm  0.03              &\hskip -0.3truecm    ...       & ...       &\hskip -0.3truecm   ...        &\hskip -0.5truecm   ...       &\hskip -0.5truecm      \\
05168+3634$^{m}$     &05 20 16.2       &+36 37 21            & 3.39(0.06)          & 2.61(0.09)          & -15.18(0.22)          &2.43(0.22)              &  & 6.08$^c$     &\hskip -0.3truecm 17.13              &\hskip -0.3truecm   7.86       &1.37       &\hskip -0.3truecm 14.79        &\hskip -0.5truecm 21.05       &\hskip -0.5truecm 1N,2N,3,5,6,7,8,10     \\
05221+4139           &05 25 39.8       &+41 41 50            & 2.16(0.06)          & 0.79(0.07)          & -25.81(0.05)    &2.75(0.13)        &  &10.36         &\hskip -0.3truecm 32.75              &\hskip -0.3truecm   6.36       &0.43       &\hskip -0.3truecm  3.05        &\hskip -0.5truecm  4.34       &\hskip -0.5truecm 2N     \\
\\
05271+3059           &05 30 21.2       &+31 01 27            & 2.32(0.07)          & 0.75(0.06)          & -19.73(0.22)    &4.36(0.22)        &  & 16.5$^b$     &\hskip -0.3truecm 72.86              &\hskip -0.3truecm   6.56       &0.37       &\hskip -0.3truecm  4.22        &\hskip -0.5truecm  6.01       &\hskip -0.5truecm 2     \\
05334+3149           &05 36 41.1       &+31 51 14            &  ...                & 1.35(0.06)          & -15.98(0.22)    &3.77(0.22)        &  & 16.5$^b$     &\hskip -0.3truecm 97.73              &\hskip -0.3truecm    ...       & ...       &\hskip -0.3truecm   ...        &\hskip -0.5truecm   ...       &\hskip -0.5truecm 2N,3N     \\
05450+0019           &05 47 34.6       &+00 20 08            & 6.87(0.09)          & 3.64(0.07)          &   9.41(0.22)    &4.19(0.22)        &  &  0.5$^d$     &\hskip -0.3truecm  0.04              &\hskip -0.3truecm  11.74       &0.73       &\hskip -0.3truecm 12.76        &\hskip -0.5truecm 18.17       &\hskip -0.5truecm      \\
06067+2138$^{m}$     &06 09 48.0       &+21 38 11            &  ...                & 3.49(0.05)          &   1.73(0.22)    &3.37(0.22)        &  & 0.61         &\hskip -0.3truecm  0.03$^{u}$        &\hskip -0.3truecm    ...       & ...       &\hskip -0.3truecm   ...        &\hskip -0.5truecm   ...       &\hskip -0.5truecm 2N,6,10     \\
06103+1523$^{m}$     &06 13 15.1       &+15 22 36            & 6.22(0.07)          & 2.97(0.09)          &  15.99(0.02)    &3.01(0.04)        &  & 3.78         &\hskip -0.3truecm  9.49              &\hskip -0.3truecm  11.04       &0.63       &\hskip -0.3truecm  6.91        &\hskip -0.5truecm  9.84       &\hskip -0.5truecm 1N,2N,3N,6,7,8     \\
\\
06104+1524A          &06 13 21.3       &+15 23 57            &  ...                & 3.24(0.19)          &  16.26(0.03)    &2.50(0.07)        &  & 3.87         &\hskip -0.3truecm 10.59              &\hskip -0.3truecm    ...       & ...       &\hskip -0.3truecm   ...        &\hskip -0.5truecm   ...       &\hskip -0.5truecm 2N,3N,5N,6     \\
06306+0232           &06 33 15.8       &+02 30 22            &  ...                & 1.43(0.08)          &  25.49(0.03)    &2.28(0.06)        &  & 3.32         &\hskip -0.3truecm  0.88              &\hskip -0.3truecm    ...       & ...       &\hskip -0.3truecm   ...        &\hskip -0.5truecm   ...       &\hskip -0.5truecm      \\
06331+1102           &06 35 56.0       &+11 00 18            &  ...                & 2.66(0.10)          &  22.71(0.22)    &2.75(0.22)        &  & 3.89         &\hskip -0.3truecm  2.58              &\hskip -0.3truecm    ...       & ...       &\hskip -0.3truecm   ...        &\hskip -0.5truecm   ...       &\hskip -0.5truecm 6     \\
06381+1039           &06 40 58.0       &+10 36 49            & 9.47(0.19)          & 2.64(0.10)          &   7.64(0.02)    &2.92(0.05)        &  & 1.11         &\hskip -0.3truecm  0.20$^{u}$        &\hskip -0.3truecm  14.50       &0.32       &\hskip -0.3truecm  4.99        &\hskip -0.5truecm  7.11       &\hskip -0.5truecm 6     \\
06382+1017           &06 41 03.3       &+10 15 01            & 7.70(0.08)          & 3.52(0.09)          &   7.60(0.01)    &2.93(0.04)        &  &  1.1         &\hskip -0.3truecm  0.17              &\hskip -0.3truecm  12.63       &0.60       &\hskip -0.3truecm  7.88        &\hskip -0.5truecm 11.22       &\hskip -0.5truecm      \\
\\
06501+0143           &06 52 45.6       &+01 40 15            &  ...                & 1.73(0.07)          &  45.07(0.02)    &2.85(0.05)        &  & 6.52         &\hskip -0.3truecm  9.44              &\hskip -0.3truecm   ...        & ...       &\hskip -0.3truecm   ...        &\hskip -0.5truecm   ...       &\hskip -0.5truecm      \\
07024$-$1102$^{m}$   &07 04 45.7       &$-$11 07 15          &  ...                & 4.47(0.06)          &  16.90(0.22)    &1.89(0.22)              &  & 1.64         &\hskip -0.3truecm  0.57              &\hskip -0.3truecm   ...        & ...       &\hskip -0.3truecm   ...        &\hskip -0.5truecm   ...       &\hskip -0.5truecm 6     \\
07111$-$1211         &07 13 29.9       &$-$12 16 51          &  ...                & 1.22(0.08)          &  15.71(0.03)    &2.37(0.09)        &  & 1.53         &\hskip -0.3truecm  0.20              &\hskip -0.3truecm   ...        & ...       &\hskip -0.3truecm   ...        &\hskip -0.5truecm   ...       &\hskip -0.5truecm      \\
07119$-$1210A        &07 14 17.7       &$-$12 15 14          &  ...                & 2.04(0.10)          &  15.23(0.03)    &2.67(0.06)        &  & 1.48         &\hskip -0.3truecm  0.23              &\hskip -0.3truecm   ...        & ...       &\hskip -0.3truecm   ...        &\hskip -0.5truecm   ...       &\hskip -0.5truecm 2N     \\
07207$-$1435         &07 23 01.3       &$-$14 41 33          &  ...                & 1.95(0.13)          &  53.17(0.04)    &2.77(0.09)        &  & 5.58         &\hskip -0.3truecm 14.50              &\hskip -0.3truecm   ...        & ...       &\hskip -0.3truecm   ...        &\hskip -0.5truecm   ...       &\hskip -0.5truecm 2       \\
\\
17576$-$1845         &18 00 34.3       &$-$18 45 17          &  ...                &1.06(0.12)           & 21.96(0.08)    &4.38(0.20)         &  & 3.11         &\hskip -0.3truecm  1.35              &\hskip -0.3truecm   ...        & ...       &\hskip -0.3truecm   ...        &\hskip -0.5truecm   ...       &\hskip -0.5truecm        \\
18145$-$1557         &18 17 26.7       &$-$15 56 20          &  ...                &1.81(0.20)           & 25.93(0.07)    &4.09(0.19)         &  & 2.81         &\hskip -0.3truecm  4.07              &\hskip -0.3truecm   ...        & ...       &\hskip -0.3truecm   ...        &\hskip -0.5truecm   ...       &\hskip -0.5truecm 3N       \\
18205$-$1316         &18 23 21.6       &$-$13 15 02          &  ...                &2.43(0.12)           & 22.48(0.22)    &2.89(0.22)         &  & 2.24         &\hskip -0.3truecm  1.52              &\hskip -0.3truecm   ...        & ...       &\hskip -0.3truecm   ...        &\hskip -0.5truecm   ...       &\hskip -0.5truecm        \\
18236$-$1241         &18 26 24.7       &$-$12 39 37          &  ...                &2.07(0.10)$^{s}$     & 64.57(0.03)    &4.33(0.08)         &  &  4.6         &\hskip -0.3truecm 15.85              &\hskip -0.3truecm   ...        & ...       &\hskip -0.3truecm   ...        &\hskip -0.5truecm   ...       &\hskip -0.5truecm        \\
18278$-$0212         &18 30 28.0       &$-$02 10 48          &  ...                &1.79(0.08)           &  5.86(0.22)    &3.01(0.22)         &  & 0.39         &\hskip -0.3truecm  0.02              &\hskip -0.3truecm   ...        & ...       &\hskip -0.3truecm   ...        &\hskip -0.5truecm   ...       &\hskip -0.5truecm        \\
\\
18301$-$0853         &18 32 55.2       &$-$08 51 23          &  ...                &2.19(0.14)           & 79.05(0.04)    &3.35(0.10)         &  & 4.97         &\hskip -0.3truecm 13.66              &\hskip -0.3truecm   ...        & ...       &\hskip -0.3truecm   ...        &\hskip -0.5truecm   ...       &\hskip -0.5truecm        \\
18314$-$0820         &18 34 09.0       &$-$08 17 52          &  ...                &4.07(0.08)           &105.17(0.22)    &4.01(0.22)         &  & 6.08         &\hskip -0.3truecm 24.52              &\hskip -0.3truecm   ...        & ...       &\hskip -0.3truecm   ...        &\hskip -0.5truecm   ...       &\hskip -0.5truecm 3N       \\
18324$-$0855         &18 35 10.5       &$-$08 52 35          &  ...                &0.92(0.10)           &  3.72(0.03)    &0.88(0.09)         &  & 0.26         &\hskip -0.3truecm  0.03              &\hskip -0.3truecm   ...        & ...       &\hskip -0.3truecm   ...        &\hskip -0.5truecm   ...       &\hskip -0.5truecm 1       \\
18358$-$0112         &18 38 25.9       &$-$01 09 51          &  ...                &2.51(0.07)           &  9.43(0.22)    &3.67(0.22)         &  & 0.67         &\hskip -0.3truecm  0.05              &\hskip -0.3truecm   ...        & ...       &\hskip -0.3truecm   ...        &\hskip -0.5truecm   ...       &\hskip -0.5truecm        \\
18403$-$0445         &18 38 34.4       &$-$03 32 06          &  ...                &1.76(0.10)$^{s}$     & 45.74(0.03)    &3.25(0.08)         &  & 3.16         &\hskip -0.3truecm  4.66              &\hskip -0.3truecm   ...        & ...       &\hskip -0.3truecm   ...        &\hskip -0.5truecm   ...       &\hskip -0.5truecm 1,6       \\
\\
18502+0033           &18 52 47.6       &+00 36 51            &  ...                &2.33(0.10)           &104.90(0.03)    &5.26(0.07)         &  &    1$^*$     &\hskip -0.3truecm  0.50              &\hskip -0.3truecm   ...        & ...       &\hskip -0.3truecm   ...        &\hskip -0.5truecm   ...       &\hskip -0.5truecm        \\
18502+0034           &18 52 50.9       &+00 38 03            &  ...                &2.40(0.12)           &104.70(0.04)    &5.55(0.09)         &  &    1$^*$     &\hskip -0.3truecm  0.52              &\hskip -0.3truecm   ...        & ...       &\hskip -0.3truecm   ...        &\hskip -0.5truecm   ...       &\hskip -0.5truecm 6       \\
18532+0047           &18 55 50.6       &+00 51 22            &  ...                &1.24(0.09)$^{s}$     & 58.94(0.05)    &4.65(0.13)         &  & 3.83         &\hskip -0.3truecm  8.96              &\hskip -0.3truecm   ...        & ...       &\hskip -0.3truecm   ...        &\hskip -0.5truecm   ...       &\hskip -0.5truecm 3N,8       \\
18545+0202           &18 57 02.6       &+02 06 23            &  ...                &1.45(0.10)$^{s}$     & 44.87(0.04)    &3.68(0.13)         &  & 2.99         &\hskip -0.3truecm  3.45              &\hskip -0.3truecm   ...        & ...       &\hskip -0.3truecm   ...        &\hskip -0.5truecm   ...       &\hskip -0.5truecm        \\
18578+0313           &19 00 21.1       &+03 17 45            &  ...                &0.97(0.08)$^{s}$     & 59.74(0.06)    &5.16(0.14)         &  & 3.93         &\hskip -0.3truecm  3.53              &\hskip -0.3truecm   ...        & ...       &\hskip -0.3truecm   ...        &\hskip -0.5truecm   ...       &\hskip -0.5truecm        \\
\\
19002+0454           &19 02 42.0       &+04 58 49            &  ...                &1.56(0.13)$^{s}$     & 68.58(0.06)    &3.75(0.15)         &  & 4.67         &\hskip -0.3truecm  8.71              &\hskip -0.3truecm   ...        & ...       &\hskip -0.3truecm   ...        &\hskip -0.5truecm   ...       &\hskip -0.5truecm 3N,7       \\
19011+0450           &19 03 36.9       &+04 55 15            &  ...                &1.38(0.13)           & 50.07(0.07)    &4.52(0.17)         &  & 3.35         &\hskip -0.3truecm  4.28              &\hskip -0.3truecm   ...        & ...       &\hskip -0.3truecm   ...        &\hskip -0.5truecm   ...       &\hskip -0.5truecm        \\
19029+0556           &19 05 23.8       &+06 01 24            &  ...                &2.11(0.13)           & 58.44(0.22)    &4.54(0.22)         &  & 3.97         &\hskip -0.3truecm 12.65              &\hskip -0.3truecm   ...        & ...       &\hskip -0.3truecm   ...        &\hskip -0.5truecm   ...       &\hskip -0.5truecm        \\
19031+0621           &19 05 36.4       &+06 26 09            &  ...                &1.65(0.10)           & 73.36(0.03)    &2.38(0.07)         &  & 5.41         &\hskip -0.3truecm 24.87              &\hskip -0.3truecm   ...        & ...       &\hskip -0.3truecm   ...        &\hskip -0.5truecm   ...       &\hskip -0.5truecm 1       \\
19056+0624           &19 08 02.9       &+06 29 11            &  ...                &1.16(0.12)$^{s}$     & 66.07(0.06)    &3.44(0.17)         &  & 4.66         &\hskip -0.3truecm  4.51              &\hskip -0.3truecm   ...        & ...       &\hskip -0.3truecm   ...        &\hskip -0.5truecm   ...       &\hskip -0.5truecm        \\
\\
19205+1358           &19 22 53.9       &+14 04 11            & 2.10(0.09)          &0.60(0.07)           & 61.86(0.07)    &3.43(0.17)         &  &    1$^*$     &\hskip -0.3truecm  0.31$^{u}$        &\hskip -0.3truecm  6.28        &0.32       &\hskip -0.3truecm  2.80        &\hskip -0.5truecm  3.99       &\hskip -0.5truecm        \\
19215+1410           &19 23 49.4       &+14 16 20            &  ...                &0.80(0.07)           & 58.00(0.07)    &5.43(0.16)         &  &    1$^*$     &\hskip -0.3truecm  0.16              &\hskip -0.3truecm   ...        & ...       &\hskip -0.3truecm   ...        &\hskip -0.5truecm   ...       &\hskip -0.5truecm 6N       \\
19216+1658           &19 23 52.4       &+17 04 01            & 3.08(0.10)          &0.92(0.07)           &  1.67(0.06)    &6.48(0.14)         &  & 10.6         &\hskip -0.3truecm 62.46$^{u}$        &\hskip -0.3truecm  7.50        &0.34       &\hskip -0.3truecm  6.18        &\hskip -0.5truecm  8.80       &\hskip -0.5truecm        \\
19282+1742           &19 30 30.5       &+17 48 30            &  ...                &2.82(0.08)           & 61.14(0.22)    &2.84(0.22)         &  &    1$^*$     &\hskip -0.3truecm  0.49              &\hskip -0.3truecm   ...        & ...       &\hskip -0.3truecm   ...        &\hskip -0.5truecm   ...       &\hskip -0.5truecm 3N,6       \\
19286+1722           &19 30 54.0       &+17 28 46            &  ...                &0.88(0.07)           & 45.48(0.05)    &3.24(0.10)         &  & 4.78         &\hskip -0.3truecm  6.57              &\hskip -0.3truecm   ...        & ...       &\hskip -0.3truecm   ...        &\hskip -0.5truecm   ...       &\hskip -0.5truecm 6       \\
\\
19291+1713           &19 31 23.4       &+17 19 42            &  ...                &1.45(0.07)$^{s}$     & 48.07(0.03)    &3.41(0.07)         &  &    1$^*$     &\hskip -0.3truecm  0.38              &\hskip -0.3truecm   ...        & ...       &\hskip -0.3truecm   ...        &\hskip -0.5truecm   ...       &\hskip -0.5truecm        \\
19298+1707           &19 32 08.5       &+17 13 35            & 4.97(0.15)$^{s}$    &0.87(0.07)           & 56.61(0.05)    &2.64(0.11)         &  &    1$^*$     &\hskip -0.3truecm  0.19$^{u}$        &\hskip -0.3truecm  9.67        &0.19       &\hskip -0.3truecm  1.69        &\hskip -0.5truecm  2.40       &\hskip -0.5truecm        \\
19348+2229           &19 36 59.8       &+22 36 08            &  ...                &0.82(0.05)$^{s}$     & 28.48(0.02)    &1.31(0.05)         &  & 2.66         &\hskip -0.3truecm  1.10              &\hskip -0.3truecm   ...        & ...       &\hskip -0.3truecm   ...        &\hskip -0.5truecm   ...       &\hskip -0.5truecm 6N       \\
19368+2239           &19 38 58.1       &+22 46 32            & 6.19(0.08)$^{s}$    &3.20(0.09)           & 36.35(0.22)    &3.46(0.22)         &  &    1$^*$     &\hskip -0.3truecm  0.33              &\hskip -0.3truecm 11.01        &0.70       &\hskip -0.3truecm  9.47        &\hskip -0.5truecm 13.48       &\hskip -0.5truecm 3N,6,7,8       \\
19406+2333           &19 42 44.4       &+23 40 30            & 3.03(0.05)$^{s}$    &1.06(0.08)           &  1.01(0.05)    &3.56(0.12)         &  & 8.74         &\hskip -0.3truecm  8.87              &\hskip -0.3truecm  7.44        &0.41       &\hskip -0.3truecm  4.09        &\hskip -0.5truecm  5.83       &\hskip -0.5truecm 6       \\
\\
19413+2349           &19 43 28.3       &+23 56 58            & 9.20(0.16)          &3.85(0.09)           & 22.33(0.01)    &1.88(0.03)         &  & 2.02         &\hskip -0.3truecm  0.60              &\hskip -0.3truecm 14.22        &0.53       &\hskip -0.3truecm  5.19        &\hskip -0.5truecm  7.39       &\hskip -0.5truecm 6       \\
19415+2312           &19 43 39.7       &+23 20 06            & 6.79(0.07)          &2.05(0.06)           & 27.53(0.02)    &5.51(0.05)         &  & 2.72         &\hskip -0.3truecm  2.03$^{u}$        &\hskip -0.3truecm 11.66        &0.35       &\hskip -0.3truecm  7.97        &\hskip -0.5truecm 11.35       &\hskip -0.5truecm 6       \\
19560+3135           &19 58 03.3       &+31 44 07            & 4.11(0.09)          &1.11(0.11)           &-65.10(0.22)    &2.19(0.22)         &  &13.29         &\hskip -0.3truecm 50.84              &\hskip -0.3truecm  8.70        &0.30       &\hskip -0.3truecm  2.08        &\hskip -0.5truecm  2.96       &\hskip -0.5truecm 3N       \\
19589+3320           &20 00 52.6       &+33 29 08            & 7.34(0.10)          &1.88(0.08)           &-22.51(0.03)    &4.78(0.07)         &  & 8.44         &\hskip -0.3truecm 37.18              &\hskip -0.3truecm 12.25        &0.29       &\hskip -0.3truecm  6.03        &\hskip -0.5truecm  8.58       &\hskip -0.5truecm 1       \\
20050+2720           &20 07 06.7       &+27 28 53            & 3.79(0.08)$^{s}$    &2.85(0.11)           &  6.39(0.02)    &4.05(0.06)         &  &  0.7$^a$     &\hskip -0.3truecm  0.26              &\hskip -0.3truecm  8.33        &1.31       &\hskip -0.3truecm 15.98        &\hskip -0.5truecm 22.75       &\hskip -0.5truecm 3,7  \\
\\
20062+3550           &20 08 09.8       &+35 59 20            & 5.09(0.07)          &2.37(0.17)           &  0.99(0.22)    &2.26(0.22)        &  & 5.24         &\hskip -0.3truecm 18.02              &\hskip -0.3truecm  9.80        &0.61       &\hskip -0.3truecm  4.74        &\hskip -0.5truecm  6.74       &\hskip -0.5truecm 1,3,6,7,8  \\
20067+3415$^{m}$     &20 08 41.3       &+34 24 19            & 4.40(0.06)          &2.64(0.06)           & 13.68(0.02)    &3.62(0.04)         &  & 2.36         &\hskip -0.3truecm  1.14              &\hskip -0.3truecm  9.03        &0.88       &\hskip -0.3truecm  9.57        &\hskip -0.5truecm 13.63       &\hskip -0.5truecm 6  \\
20094+2744           &20 11 29.1       &+27 53 16            &  ...                &2.91(0.14)           & 12.10(0.03)    &2.59(0.06)         &  & 1.11         &\hskip -0.3truecm  0.12              &\hskip -0.3truecm   ...        & ...       &\hskip -0.3truecm   ...        &\hskip -0.5truecm   ...       &\hskip -0.5truecm   \\
20103+3633           &20 12 13.9       &+36 42 59            & 2.53(0.09)          &1.70(0.24)           &-36.23(0.07)    &2.65(0.18)         &  & 8.91         &\hskip -0.3truecm 29.26              &\hskip -0.3truecm  6.83        &1.04       &\hskip -0.3truecm  7.32        &\hskip -0.5truecm 10.42       &\hskip -0.5truecm 3N  \\
20116+3605           &20 13 33.6       &+36 14 55            & 3.38(0.10)$^{s}$    &1.15(0.07)$^{s}$     &-53.07(0.03)    &2.50(0.07)         &  &10.69         &\hskip -0.3truecm 72.70              &\hskip -0.3truecm  7.85        &0.40       &\hskip -0.3truecm  2.89        &\hskip -0.5truecm  4.11       &\hskip -0.5truecm   \\
\\
20145+3645           &20 16 27.5       &+36 54 58            & 2.70(0.11)          &1.10(0.07)           &-56.37(0.04)    &3.54(0.12)         &  &10.86         &\hskip -0.3truecm 50.37              &\hskip -0.3truecm  7.04        &0.50       &\hskip -0.3truecm  4.76        &\hskip -0.5truecm  6.78       &\hskip -0.5truecm   \\
20149+3913$^{m}$     &20 16 42.6       &+39 23 15            & 6.91(0.08)          &4.61(0.13)           &  3.82(0.02)    &2.97(0.04)         &  &1.7$^f$         &\hskip -0.3truecm  0.30$^{u}$        &\hskip -0.3truecm 11.79        &1.06       &\hskip -0.3truecm 16.85        &\hskip -0.5truecm 24.00       &\hskip -0.5truecm 6,9  \\
20178+3723           &20 19 43.1       &+37 33 13            & 6.03(0.09)          &1.96(0.10)           &  3.93(0.02)    &1.84(0.05)         &  & 4.08         &\hskip -0.3truecm  1.02$^{u}$        &\hskip -0.3truecm 10.83        &0.38       &\hskip -0.3truecm  2.69        &\hskip -0.5truecm  3.83       &\hskip -0.5truecm   \\
20227+4154           &20 24 31.4       &+42 04 17            & 7.74(0.08)          &3.18(0.12)           &  5.54(0.02)    &2.72(0.05)         &  &    2$^a$     &\hskip -0.3truecm  2.74              &\hskip -0.3truecm 12.68        &0.52       &\hskip -0.3truecm  6.37        &\hskip -0.5truecm  9.06       &\hskip -0.5truecm 3,7  \\
20231+3440           &20 25 16.0       &+34 50 06            & 3.60(0.07)          &1.90(0.14)           &  5.46(0.04)    &2.70(0.09)         &  &    1$^a$     &\hskip -0.3truecm  0.19              &\hskip -0.3truecm  8.11        &0.72       &\hskip -0.3truecm  5.72        &\hskip -0.5truecm  8.14       &\hskip -0.5truecm   \\
\\
20261+3922           &20 27 58.8       &+39 32 07            & 2.63(0.08)$^{s}$    &1.19(0.19)           &-54.99(0.09)    &2.79(0.20)         &  &10.01         &\hskip -0.3truecm 32.70              &\hskip -0.3truecm  6.95        &0.57       &\hskip -0.3truecm  4.28        &\hskip -0.5truecm  6.09       &\hskip -0.5truecm   \\
20281+4038           &20 29 54.8       &+40 48 52            & 5.50(0.08)$^{s}$    &2.05(0.06)           & -2.51(0.02)    &2.36(0.04)         &  & 4.11         &\hskip -0.3truecm  6.56              &\hskip -0.3truecm 10.25        &0.45       &\hskip -0.3truecm  3.86        &\hskip -0.5truecm  5.50       &\hskip -0.5truecm   \\
20290+4052           &20 30 50.8       &+41 02 25            & 3.70(0.08)$^{s}$    &1.60(0.10)           & -1.51(0.04)    &3.15(0.10)         &  & 3.89         &\hskip -0.3truecm  2.79              &\hskip -0.3truecm  8.23        &0.54       &\hskip -0.3truecm  5.12        &\hskip -0.5truecm  7.28       &\hskip -0.5truecm   \\
20326+3757           &20 34 33.0       &+38 08 02            & 9.05(0.11)          &2.99(0.09)           &  2.79(0.22)    &5.81(0.22)         &  & 3.66         &\hskip -0.3truecm  2.74$^{u}$        &\hskip -0.3truecm 14.06        &0.39       &\hskip -0.3truecm 11.72        &\hskip -0.5truecm 16.69       &\hskip -0.5truecm   \\
20330+4109           &20 34 48.5       &+41 20 21            & 4.07(0.11)$^{s}$    &0.67(0.07)$^{s}$     &-30.82(0.05)    &2.19(0.12)         &  & 7.09         &\hskip -0.3truecm 10.04$^{u}$        &\hskip -0.3truecm  8.65        &0.17       &\hskip -0.3truecm  1.19        &\hskip -0.5truecm  1.69       &\hskip -0.5truecm   \\
\\
20345+4024           &20 36 24.1       &+40 35 08            & 4.12(0.10)          &0.82(0.09)           &  1.31(0.06)    &2.58(0.13)         &  & 3.34         &\hskip -0.3truecm  1.59$^{u}$        &\hskip -0.3truecm  8.71        &0.21       &\hskip -0.3truecm  1.73        &\hskip -0.5truecm  2.47       &\hskip -0.5truecm   \\
20444+4629           &20 46 08.3       &+46 40 41            & 6.51(0.09)          &3.65(0.07)           & -3.71(0.01)    &2.09(0.02)         &  & 2.91         &\hskip -0.3truecm  3.98              &\hskip -0.3truecm 11.36        &0.80       &\hskip -0.3truecm  6.67        &\hskip -0.5truecm  9.50       &\hskip -0.5truecm 2N,3N,7  \\
20508+4825           &20 52 28.2       &+48 36 30            & 1.55(0.05)$^{s}$    &0.92(0.07)           & -6.63(0.04)    &2.58(0.11)         &  & 2.98         &\hskip -0.3truecm  1.17$^{u}$        &\hskip -0.3truecm  5.55        &0.84       &\hskip -0.3truecm  5.27        &\hskip -0.5truecm  7.51       &\hskip -0.5truecm 2N  \\
21025+4912           &21 04 15.4       &+49 24 25            & 1.32(0.04)$^{s}$    &0.74(0.06)           &-73.44(0.05)    &2.86(0.12)         &  &10.08         &\hskip -0.3truecm  9.15$^{u}$        &\hskip -0.3truecm  5.23        &0.76       &\hskip -0.3truecm  5.27        &\hskip -0.5truecm  7.50       &\hskip -0.5truecm 2  \\
21246+5512           &21 26 14.4       &+55 25 57            & 0.76(0.05)          &0.49(0.06)           &-69.33(0.05)    &1.84(0.13)         &  & 8.72         &\hskip -0.3truecm 15.33              &\hskip -0.3truecm  4.37        &0.93       &\hskip -0.3truecm  4.14        &\hskip -0.5truecm  5.90       &\hskip -0.5truecm 2N  \\
\\
21293+5535           &21 30 55.7       &+55 48 49            & 4.42(0.06)          &1.17(0.06)           &-71.13(0.03)    &4.13(0.09)         &  & 8.85         &\hskip -0.3truecm  8.25$^{u}$        &\hskip -0.3truecm  9.05        &0.30       &\hskip -0.3truecm  3.97        &\hskip -0.5truecm  5.65       &\hskip -0.5truecm 2,4  \\
21334+5039           &21 35 09.2       &+50 53 09            & 4.64(0.07)          &2.36(0.07)           &-45.05(0.22)    &2.75(0.22)         &  &    5$^a$     &\hskip -0.3truecm 23.53              &\hskip -0.3truecm  9.30        &0.68       &\hskip -0.3truecm  6.21        &\hskip -0.5truecm  8.84       &\hskip -0.5truecm 2  \\
21379+5106           &21 39 40.8       &+51 20 35            & 2.88(0.07)          &1.39(0.07)           &-42.27(0.22)    &2.17(0.22)          &  &  5.9         &\hskip -0.3truecm  5.18              &\hskip -0.3truecm  7.25        &0.63       &\hskip -0.3truecm  3.73        &\hskip -0.5truecm  5.32       &\hskip -0.5truecm 2  \\
21391+5026           &21 40 57.3       &+50 39 53            & 2.56(0.07)          &1.02(0.07)           &-40.55(0.04)    &2.50(0.09)         &  & 5.77         &\hskip -0.3truecm  5.67              &\hskip -0.3truecm  6.86        &0.48       &\hskip -0.3truecm  3.22        &\hskip -0.5truecm  4.58       &\hskip -0.5truecm 2  \\
21391+5802$^{m}$     &21 40 42.4       &+58 16 10            &  ...                &5.80(0.07)           &  0.68(0.01)    &2.99(0.02)         &  & 0.75$^e$     &\hskip -0.3truecm  0.26              &\hskip -0.3truecm   ...        & ...       &\hskip -0.3truecm   ...        &\hskip -0.5truecm   ...       &\hskip -0.5truecm 1N,2N,3,7,8  \\
\\
21418+5403           &21 43 29.8       &+54 16 56            & 3.62(0.06)          &0.86(0.06)           &-60.07(0.04)    &3.24(0.10)         &  & 7.51         &\hskip -0.3truecm 22.24              &\hskip -0.3truecm  8.13        &0.26       &\hskip -0.3truecm  2.51        &\hskip -0.5truecm  3.58       &\hskip -0.5truecm 2N  \\
21519+5613           &21 53 39.2       &+56 27 46            & 4.26(0.06)          &2.07(0.08)           &-62.72(0.02)    &3.11(0.05)         &  &  7.3$^a$     &\hskip -0.3truecm 15.75              &\hskip -0.3truecm  8.87        &0.64       &\hskip -0.3truecm  6.31        &\hskip -0.5truecm  8.98       &\hskip -0.5truecm 2,3,4,8  \\
22051+5848           &22 06 50.7       &+59 02 47            & 2.40(0.06)          &1.51(0.07)           & -1.77(0.02)    &1.39(0.04)         &  & 0.77         &\hskip -0.3truecm  0.08$^{u}$        &\hskip -0.3truecm  6.66        &0.93       &\hskip -0.3truecm  3.38        &\hskip -0.5truecm  4.82       &\hskip -0.5truecm 2N,4  \\
22198+6336$^{m}$     &22 21 27.6       &+63 51 42            &10.55(0.01)          &5.39(0.07)           &-11.10(0.22)     &2.65(0.22)          &  & 1.67         &\hskip -0.3truecm  1.53$^{u}$        &\hskip -0.3truecm 15.63        &0.70       &\hskip -0.3truecm 11.49        &\hskip -0.5truecm 16.36       &\hskip -0.5truecm 1N,2N,3,7,8  \\
22305+5803           &22 32 24.3       &+58 18 58            & 4.57(0.07)          &1.80(0.06)           &-52.41(0.02)    &2.68(0.05)         &  & 5.94         &\hskip -0.3truecm 14.28              &\hskip -0.3truecm  9.22        &0.48       &\hskip -0.3truecm  4.25        &\hskip -0.5truecm  6.05       &\hskip -0.5truecm 2,4,6,8  \\
\\
22506+5944$^{m}$     &22 52 38.6       &+60 00 56            &11.01(0.07)          &3.18(0.05)           &-51.61(0.22)     &2.43(0.22)          &  &  3.5$^d$     &\hskip -0.3truecm  6.83              &\hskip -0.3truecm 16.11        &0.34       &\hskip -0.3truecm  5.56        &\hskip -0.5truecm  7.92       &\hskip -0.5truecm 1N,2N,3,4N,6,7,8,10  \\
22539+5758           &22 56 00.0       &+58 14 46            & 7.32(0.06)          &3.59(0.07)           &-54.19(0.01)    &2.97(0.03)         &  &  3.5$^d$     &\hskip -0.3truecm 10.68              &\hskip -0.3truecm 12.23        &0.66       &\hskip -0.3truecm  8.47        &\hskip -0.5truecm 12.07       &\hskip -0.5truecm 2,4,5  \\
23011+6126           &23 03 13.1       &+61 42 26            & 5.12(0.07)          &2.25(0.06)           &-11.03(0.22)    &1.93(0.22)         &  & 0.73$^d$     &\hskip -0.3truecm  0.09              &\hskip -0.3truecm  9.83        &0.56       &\hskip -0.3truecm  3.75        &\hskip -0.5truecm  5.34       &\hskip -0.5truecm 2,4,6  \\
\hline
\end{longtable}
Note.---Column (1) is source name shown in order of ascending
\textit{IRAS} number. Columns (2) and (3) list J2000 equatorial
coordinates of each source. Column (4) lists the main beam
temperature of $^{12}$CO~\textit{J}\,=\,2\,--\,1 if available, and
Cols. (5)--(7) list the main beam temperature, local standard of
rest velocity and line width of $^{13}$CO~\textit{J}\,=\,2\,--\,1,
with $1\sigma$ rms level in parenthesis. Column (8) is kinematic
distance except for otherwise labeled. Column (9) lists the
bolometric luminosity with asterisk (*) if one (or two)
\textit{IRAS} band flux is upper limit. Column (10) is excitation
temperature. Column (11) is the opacity of ${}^{13}$CO. Columns
(12)--(13) present average ${}^{13}$CO and H$_{2}$ column densities.
Column (14) lists former works if available.

References.--- (1) 6.7\,GHz methanol maser survey
\citep{2000A&AS..143..269S}; (2) 22\,GHz water maser survey
\citep{1993A&AS...98..589W}; (3) 22\,GHz water maser
\citep{1991A&A...246..249P}; (4) 1665/67\,MHz OH maser survey
\citep{1993A&AS...98..589W}; (5) HCO$^+$~\textit{J}\,=\,1\,--\,0
survey \citep{1987MNRAS.228...43R}; (6) CS~\textit{J}\,=\,2\,--\,1
survey \citep{1996A&AS..115...81B}; (7)
NH$_3$~(\textit{J,K})\,=\,(1,1) (2,2) survey
\citep{1996A&A...308..573M}; (8) CO~\textit{J}\,=\,2\,--\,1 outflow
mapping \citep{2005ApJ...625..864Z, 2006ApJ...643..978K}; (9) CO,
CS mapping \citep{2004AJ....128.1716A}; (10) CO mapping
\citep{2008MNRAS.391..869G}; (11) CO mapping
\citep{2003NewA....8..191L}. N means
non-detection. \\
\\
$^{\rm a-f}${~distance from $^{\rm a}$ \cite{2004A&A...426..503W},
$^{\rm b}$ \cite{1993A&AS...98..589W}, $^{\rm c}$
\cite{1996A&A...308..573M}, $^{\rm d}$ \cite{1993A&AS...98...51H},
$^{\rm e}$ \cite{1979A&A....75..345M},
$^{\rm f}$ \cite{2007A&A...476.1243M}};\\
$^{\rm *}${~kinematic distance unavailable, set as 1\,kpc;}\\
$^{\rm m}${~mapped source;}\\
$^{\rm s}${~more than one velocity component, only the
strongest one listed;}\\
$^{\rm u}${~upper limit.}

\end{landscape}
} %----------------------------------END online table.-----------------------------

\end{document}